\documentclass[journal=jacsat,manuscript=article]{achemso}
\usepackage{chemformula} 
\usepackage[T1]{fontenc} 
\usepackage{graphicx}
\usepackage{xcolor}
\usepackage{amssymb}

\author{John F. Dobson} 
\affiliation{School of Environment and Science, Griffith University, Nathan Queensland 4111, Australia}
\email{j.dobson@griffith.edu.au}
\author{Alberto Ambrosetti}
\affiliation{Dipartimento di Fisica e Astronomia, Universit\`a degli Studi di Padova, via Marzolo 8, 35131, Padova, Italy}

\title{MBD+C: how to incorporate  metallic character
 into atom-based dispersion energy schemes}


\begin{document}

\tableofcontents

\begin{abstract}
The dispersion component of the van der Waals (vdW)  interaction in low-dimensional metals is known to \ exhibit
anomalous ''Type -C non-additivity'' [Int. J. Quantum Chem. 114, 1157 (2014)]. This causes dispersion energy behavior,  at asymptotically large separations, that is missed by popular atom-based schemes for
dispersion energy calculations.   For example, the dispersion interaction energy between parallel metallic nanotubes at separation D falls off  aymptotically as approximately  $D^{-2}$, whereas  current atom-based  schemes predict $D^{-5}$  asymptotically.    To date it has not been clear whether current atom-based  theories also give the dispersion interaction inaccurately at smaller separations  for low- dimensional metals.
 Here we introduce a new theory that we term "MBD+C" . It permits inclusion of Type C effects efficiently within atom-based dispersion energy schemes such as Many Body Dispersion (MBD) and Universal MBD (uMBD). This allows us to investigate  asymptotic, intermediate and near-contact regimes with equal accuracy. (The large contact energy of intimate metallic bonding is not primarily governed by 
dispersion energy and is described well by semi-local density functional theory.)
Here we  apply a simplified version,"nn-MBD+C",  of our new  theory to calculate the dispersion interaction for three low-dimensional metallic systems:   parallel metallic chains of gold atoms, parallel Li-doped graphene sheets; and parallel (4,4)  armchair carbon nanotubes.  In addition to giving the correct asymptotic behavior, the new theory seamlessly  gives the dispersion energy down to near-contact  geometry, where it is similar to MBD but can give up to 15\% more dispersion energy than current MBD schemes, in the systems studied so far. This percentage increases with separation until nn-MBD+C dominates MBD at asymptotic separations.
\end{abstract}


\section{Introduction}

Discrete atom-based computational approaches have recently been popular  for calculation of  the  dispersion component of the  van der Waals (vdW) interaction. These approaches include the  "many-body dispersion" (MBD \cite{mbd,mbdrs,proof}), "fractionally ionic MBD (FI-MBD,  \cite{Gould_FIMBD_2016})" and "universal MBD " (uMBD  \cite{Kim_uMBD_2020}) schemes. \

These approaches are highly efficient numerically, and obtain
surprisingly good dispersion energy predictions  by using a lumped correlation energy
approach that avoids expensive wavefunction-based calculations. In these
schemes  each atom labelled $I$ has a dipolar polarizability 
${\bf \alpha }_{I}$ treated as that of a simple harmonic oscillator, and is allowed to
have a dynamic electric dipole moment $\vec{p}_{I}\exp (-i\omega t)$. \
These dynamic dipoles couple via the Coulomb interaction and the
resulting coupled mode frequencies $\omega_{j}$ can be calculated. \ The
dispersion interaction is contained in the sum of zero-point mode energies $%
\sum_{j}\hbar \omega_{j}/2$ of these collective modes. 
Because of the many-body treatment of the dipole-dipole coupling, 
collective modes can exhibit highly non-local correlation\cite{science}. 
But the inbuilt harmonic confinement implies that the individual atomic charges cannot 
move far away from their ionic sites.  \ \ 

\ In a metal, however, fields impinging on an atom can not only create a
dipole within the atom but can also move electrons onto other atoms. In low-dimensional metals this gives rise to gapless plasmon excitations that are not captured by MBD theory. Coupling between these plasmons
results in unusual power-law decays of the dispersion interaction between
low-dimensional metals\cite{dobson-white-rubio}, sometimes termed "Type-C'' dispersion
interactions\cite{dobson-abc,dobson-es,dobson-book}.    Similar anomalous dispersion interactions
have  been proposed in a variety of other cases with a small or zero electronic energy gap 
\cite{vdWPointingWiresAWhiteJFDPRB09,HChainsLiuAngyanDobsonJCP2011,%
LebeguePRL:2010,Gr_Cleavage_exfol,SpookyCorrelations,Nonadd_vdW_Long_Mol_surf_expt}. 
The Type-C interactions are not captured by
standard MBD calculations\cite{mbdrs,science,Kim_uMBD_2020}. 

The essential new  feature added here is
to include such charge mobility processes into an atom-based
localised-oscillator scheme for the correlation energy, such as MBD. 
In the context of electrostatics such charge transfer processes have been described previously and are important in  applications such as field emission from a nanotube \cite{Mayer_monopole_dipole_APL_05}. 
Here we are calculating electron correlation energies and this requires a dynamical description of the charge transfer process in order to determine the normal modes of oscillation.   This is achieved by introducing a discrete form of the Kohn-Sham electron  density response function,  $\chi _{0IJ}(\omega)$, that  specifies the time-varying  charge induced on atom $I$ when a time-varying potential is applied to atom $J$.  This response is then used in self-consistent linear equations linking dynamic  atomic charge and dipoles via their Coulomb interaction.  These equations yield normal modes, and  the modes' zero-point energies are summed to yield the  zero-temperature correlation energy, which contains the dispersion interaction.  If $\chi_0$ is set to zero, the new theory reverts to regular MBD.

The paper is organized as follows. We first describe the new MBD+C scheme in its most general form. Then we introduce a simplified near-neighbor approximation ("nn-MBD+C")
for the atom-based charge response $\chi _{0}$, which is them used for the remainder of the paper. Despite its simplicity, our model correctly describes the long-wavelength gapless metallic plasmon modes  that are responsible for the anomalous type-C dispersion interactions. 
 To demonstrate this approach quantitatively, we first  set up our method for a chain of atoms,
discussing stability of the collective modes, which can be problematic when describing metals by
MBD-type theories. Then we consider two parallel chains. We first solve a
strictly one-dimensional model wherein each chain contains metal atoms that
are polarizable only in the direction along the chain. We exhibit analytic
solutions for the frequencies of the combined plasmon-polarization waves of
a parallel pair of chains. \ We sum their zero-point energies to obtain the
dispersion interaction between the chains, verifying  that the correct Type-C asymptotics are achieved. 
 Then we apply the nn-MBD+C approach numerically to some more
realistic systems  that are known to exhibit type-C anomalous dispersion interactions at asymptotically large separations:
\begin{enumerate}
\item{ two parallel monoatomic chains of equally spaced gold atoms} 
\item{ two parallel sheets of metallic lithium- doped graphene}
\item{two parallel conducting carbon nanotubes}
\end{enumerate}
Our method gives predictions of the dispersion interaction in a seamless fashion at all separations from the distant asymptotic regime where it agrees with known analytic Type-C decay exponents, down to near-contact separations where the effects of metallicity were  not well known till now. 
 From the above examples we find that the nn- MBD+C dispersion energy near to contact is modestly greater than the
pure MBD prediction, but can exceed it  by 15\% in the above systems . This percentage increases with separation, with a gradual
transition to a type-C \cite{dobson-abc} metallic dispersion interaction at larger separations.
A final Discussion section summarises the new method and the results it has yielded so far, plus the prospects  for future work.
\section{General form of the MBD+C approach}
 In genera the flow of electronic charge between atoms will be described by a discrete atom-based form of the dynamic independent-electron density (charge)
response: $\chi _{0}\left( \vec{r},\vec{r}^{\prime },\omega \right)
\rightarrow \chi _{0IJ}\left( \omega \right) $  
where    $I$ and $J$ label %
''atoms'' located at positions $\vec{R}_{I}$, $\vec{R}_{J}$
Thus if small electron potential energy perturbations
$v_J \exp(-i\omega t)$  are applied at sites $J$ then the expected number of additional 
electrons induced at site $I$ is 
\begin{equation}
n_I(t) = \exp(-i\omega t) \sum_J \chi_{0IJ}(\omega) v_J
\label{Action_of_Chi0IJ}
\end{equation}
  
Assuming that we know the  independent-electron charge response $\chi_{0IJ}(\omega)$ and the atomic polarizability tensors $\boldsymbol{\alpha}_I(\omega)$ we can write a set of self-consistent (RPA) equations for  the  dynamic charges $-en_I  \exp(-i\omega t)$ and dipoles $\vec{p}_I exp(-i\omega t)$:
\begin{eqnarray}
\vec{p}_I &=& \sum_J \boldsymbol{\alpha}_I \cdot   ( \mathbf{T}_{IJ} . \vec{p}_J + \vec{u}_{IJ}n_J) 
\label{Gen_MBD+C_for_p}\\
	n_I &=& \sum_J \chi_{0IJ} (\vec{u}_{IJ} \cdot  \vec{p}_J + w_{IJ} n_J )
\label{Gen_MBD+C_for_n}
\end{eqnarray}
Here $\mathbf{T}$, $\vec{u}$ and $w$ are the dipole-dipole, dipole-charge and charge-charge Coulomb interactions respectively. The above equations constitute a set of $4N\times 4N$ homogeneous linear equations for the charges and dipoles, where $N$ is the number of atoms.   If we turn off the charge response $\chi _{0JJ}$ (appropriate for non-metallic systems)  we recover the usual $3N\times3N$ equations of MBD theory. 
The general MBD+C scheme is then implemented by finding the dipole-charge mode frequencies $\{ \omega_j  \}$ that make the determinant of the homogeneous linear equations (\ref{Gen_MBD+C_for_p}) and (\ref{Gen_MBD+C_for_n}) vanish.  The MBD+C dispersion energy at T=0K between bodies separated by distance $D$ is the separation-dependent part of the sum of mode zero-point energies $\hbar \omega /2$:
\begin{equation}
E_{disp} =\frac{\hbar}{2}\sum_j(\omega_j(D) -\omega_j(\infty))
\label{EvdW_from_mode_frequs}
\end{equation}
Equations (\ref{Gen_MBD+C_for_p}),  (\ref{Gen_MBD+C_for_n}) and (\ref{EvdW_from_mode_frequs}) define the general MBD+C scheme. This scheme requires input of the response function $\chi_{0IJ}$ , which embodies much of the new physics here. 

\section{nn-MBD+C: a simple near-neighbor version of the new MBD+C theory}
Metallic behavior is manifested as a zero-frequency divergence of the bare (independent-electron,
Kohn-Sham) microscopic linear density-density response function
 $\chi _{0}$ $\left( \vec{r},\vec{r}\;\prime,\omega \right) $. In general $\chi _{0}$ gives the
electron number density perturbation $n\left( \vec{r}\right) \exp \left(
-i\omega t\right) $ at position $\vec{r}\;$ due to the total effective
potential perturbation $v\left( \vec{r}\,^{\prime }\right) \exp \left(
-i\omega t\right) $ applied at position $\vec{r}\,^{\prime }$. Thus $n=\chi
_{0}\ast v$ where * indicates spatial convolution. \ Via quantum
perturbation theory, \ $\chi _{0}$ can be calculated from the global Kohn-Sham
orbitals of the system at hand. \ By representing the KS orbitals of a
multi-atom system in terms of localized orbitals, one can obtain   
 an expression for $\chi _{0}$ involving discrete atomic orbitals. 
  This expression is somewhat complicated, including 4-orbital terms in which
orbitals can be localized on the same atom or different atoms.
Spatial integration over atomic sites then leads to a atom-based response of the form proposed
 in Eq. (\ref{Action_of_Chi0IJ}), namely $\chi_{0IJ}(\omega)$.  
In large metallic systems,
 as suggested by electron gas  theory, $\chi _{0IJ}$ is expected to show an oscillatory, algebraically-decaying tail as 
$|\vec{R}_{I}-\vec{R}_{J}| \rightarrow \infty $. Nevertheless we will now show that a simple
nearest-neighbor Ansatz for $\chi _{0IJ}$ yields  the correct gapless dispersion
\ relation for long-wavelength plasmons, and this is the essential physics 
leading to Type-C dispersion interactions between low-dimensional metals \cite{dobson-white-rubio}. \
We stress, though, that this Ansatz is not essential for a tractable theory,
and our future work s will obtain the full $\chi _{0IJ}$  from
 a tight  binding type of approach.

In a metal, charge can move dynamically between atoms. Thus an electronic
potential energy $v_{J}=-\left| e\right| \phi _{J}$ applied to an atom
labelled $J$ in a metal will cause an increase/decrease 
 $q_{J}=-\left|e\right| n_{J}$ in the charge on that atom, while altering the charges on
other atoms to conserve total electronic charge. \ Here we simplify the
problem by restricting this charge-displacement effect to neighboring atoms.

For example, on a one-dimensional chain of identical atoms with equal
spacing $R$ we choose the following charge-conserving near-neighbor Ansatz:%
\begin{equation}
\chi _{0IJ}^{Ansatz}\equiv \chi _{0}\left( I-J\right) =B\omega ^{-2}\left(
\delta _{IJ}-\frac{1}{2}\delta _{I,J+1}-\frac{1}{2}\delta _{I,J-1}\right)
\label{Chi0Ansatz)_1Dchain}
\end{equation}%
where the constant $B$ will be chosen \ below to match the known
long-wavelength response function of a one-dimensional(1D) metal near to its
1D plasma frequency.

As a further example,  for a doped graphene layer we could choose

\begin{equation}
\chi_{0IJ}^{Ansatz}=B\omega ^{-2}\left( \delta _{IJ}-\frac{1}{3}\delta
_{IK}-\frac{1}{3}\delta _{I,L}-\frac{1}{3}\delta _{IM}\right)
\label{Chi0Ansatz_graphene}
\end{equation}%
where $K,$ $L$ and $M$ \ label the  three carbon atoms adjacent to atom 
$J $.\   $B$ is now chosen to reproduce the long-wavelength response of a 2D
metal.

More generally the nearest neighbor Ansatz for independent electron response is
\begin{eqnarray}
n_{I} &=&\sum_{J}\chi _{0IJ}^{Ansatz}v_{J}  \nonumber \\
\chi _{0IJ}^{Ansatz}
 & =&B\omega^{-2}(\delta _{IJ}+         f_{IJ})  
\label{GeneralNN_Chi0Ansatz}
\end{eqnarray}%
where $f_{IJ}=0$ unless $I$ and $J$ are nearest neighbors. 
 Charge conservation requires the weights $f_{IJ}$ to satisfy 
\begin{equation}
1+\sum_{I}f_{IJ}=0  \label{ChargeConservation}
\end{equation}

For example, on a uniform 1D chain we have $f_{IJ}=-\Delta_{IJ}/2$ to reproduce (\ref{Chi0Ansatz)_1Dchain}) and for graphene $f_{IJ}=-\Delta_{IJ}/3$ to reproduce 
(\ref{Chi0Ansatz_graphene}). Here 
 $\Delta_{IJ}$ equals $1$ when  $I$ and $J$ are  nearest
neighbors, and $0$ otherwise.  In less symmetric systems the weights are not necessarily equal, in contrast to these two simple examples.

\ The factor $\omega^{-2}$ in (\ref{GeneralNN_Chi0Ansatz}) ensures
agreement with band theory for metals, \ as shown below.

Type-C metallic effects in the dispersion energy occur most strongly
\cite{dobson-white-rubio,dobson-book} in infinite periodic systems of low spatial dimension - i.e. low-dimensional metals. For infinite low-d metals $\chi_{0}$ is most
conveniently expressed in Fourier transform ($q)$ space: 
$\chi_0(G,G',q,\omega)$. Here the $G$'s are reciprocal lattice vectors.  The $G=G'=0$ form
 is relevant here because it relates to total charge in a unit cell. It can be calculated from Bloch band theory and has the following form
 for a  metal:
\[
\chi _{0}\left( q\rightarrow 0,\omega \right)
\equiv \chi _{0}\left(0,0, q\rightarrow 0,\omega \right)
 =Aq^{2}/\omega ^{2} 
\]%

which applies for small wavenumbers $q$ and frequencies $\omega $ that are
''high'' in the sense that $\left| \omega \right| >v_{F}q$, where $v_{F\text{ 
}}$is the Fermi velocity. \ The plasmons of extended systems in $1$, 2 or 3
dimensions occur in this ''high-frequency'' regime. \ Recovery of these
plasmon modes is essential for description of Type-C dispersion physics.

For example, in a one-dimensional metal such as an intrinsically metallic
nanotube or a gold atom chain  of length $L\rightarrow \infty $ we have (see Supporting Information)
\begin{equation}
\chi _{0}^{1D}\left( q,\omega \right) \equiv \frac{1}{L}\int drdr^{\prime
}\chi _{0}\left( r,r^{\prime }\right) e^{-iq\left( r-r^{\prime }\right)
}\approx \frac{N_sN_vv_{F}}{\pi \hbar }\frac{q^{2}}{\omega ^{2}}%
\;\;as\,q\rightarrow 0  \label{Chi0_1Dmetal}
\end{equation}%
and for an isotropic  2D metal such as doped graphene we have%
\begin{equation}
\chi _{0}^{2D}\left( \vec{q},\omega \right) \approx 
\frac{N_sN_v v_{F}k_{F}}{4\pi\hbar }{}\frac{q^{2}}{\omega ^{2}}\;\;as\,q\rightarrow 0
\label{Chi0_2Dmetal}
\end{equation}

Here $N_s$ is the number of allowed electron spin orientations (usually 2 in non-magnetized metals)  and  $N_v$ is the number of "valleys" (minima) in the Bloch bandstructure diagram ($\varepsilon(k)$ vs. $k$) where a pocket of electrons can occur in the groundstate.  

In (\ref{Chi0_1Dmetal}) and (\ref{Chi0_2Dmetal}) 
  $v_{F}\equiv  {\partial \varepsilon _{k}/\partial k|_{k_{F}}}\approx
10^{6}\;m/s$ is the  characteristic  band velocity and $k_{F}$ is the
Fermi wavenumber of the 2D metal, which is dependent on the level of
doping for graphene. \ For a regular 2D metal, both the Fermi velocity $v_{F}$ and the
Fermi wavenumber $k_{F\text{ }}$ depend on the doping, and the product $%
v_{F}k_{F}$ is proportional to the areal density of metallic electrons.

In order to make the present simple atom-based description agree with these
known $q\rightarrow 0$ results for low-dimensional metallic response, we need
to Fourier-transform the real-space Ansatz (\ref{GeneralNN_Chi0Ansatz}), (%
\ref{Chi0Ansatz)_1Dchain}), (\ref{Chi0Ansatz_graphene}). \ The discrete
Fourier transform of a general translationally invariant lattice function $F%
\left[ R_{I}-R_{J}\right] $ on a periodic array of atoms at positions $%
\left\{ R_{I}\right\} $ is defined as a lattice sum 
\[
F\left( \vec{q}\right) \equiv \sum_{I}e^{-i\vec{q}\cdot\vec{R}_{I}}F\left[ I\right] 
\]%

For a periodic system the near-neighbor weights $f$ in (\ref{GeneralNN_Chi0Ansatz}) can be written 
$f_{IJ}=f(K)$ where $K$ labels the near-neighbor vectors $\vec{R}_I - \vec{R}_J = \vec{X}_K$.
Using (\ref{GeneralNN_Chi0Ansatz}) we obtain 
\begin{eqnarray*}
\chi _{0}^{Ansatz}\left[ \vec{q},\omega \right] &\equiv
&\sum_{I}e^{-i\vec{q}\cdot\vec{R}_{I}}\chi _{0}\left[ I\right] 
=B\omega ^{-2}\left(1+\sum_K f(K) e^{i\vec{q}\cdot\vec{X}_{K}}\right) \\
&\approx &B\omega ^{-2}\left( \left( 1+\sum_K f(K)\right)
 +i\vec{q} \cdot\sum_K f(K)\vec{X}_{K} + O\left( q^{2}\right) \right)
\end{eqnarray*}%

 The first bracket on the right side is zero
by charge conservation (Eq (\ref{ChargeConservation})). \ If there is high
symmetry (as for example in (\ref{Chi0Ansatz)_1Dchain}) or  
(\ref{Chi0Ansatz_graphene}) ) so that $\sum_{K} f(K)\vec{X}_{K}=\vec{0}$ then $%
\chi _{0}\left( q,\omega \right) $ is quadratic in $q$ for small $q$, which
is the correct behavior for a metal. If there is lower symmetry, then the
nearest-neighbor weights $f(K)$ should be chosen to vanish
the term that is linear in $q$. That is, we should choose the $f(K) $ such that%
\[
\sum_K f(K) \vec{X}_{K}=\vec{0} 
\]

\ The discrete Fourier transform of the 1D chain response (\ref%
{Chi0Ansatz)_1Dchain}) is%
\begin{eqnarray}
\chi _{0}^{ansatz}\left[ q,\omega \right] &\equiv &\sum_{m=-\infty }^{\infty
}e^{-iqmR}\chi _{0}\left( m\right) =\frac{B}{\omega ^{2}}\left( 1-\frac{1}{2}%
e^{iqR}-\frac{1}{2}e^{-iqR}\right) \\
&=&\frac{B}{\omega ^{2}}\left( 1-\cos \left( qR\right) \right) =\frac{B}{%
\omega ^{2}}2\sin ^{2}\left( \frac{qR}{2}\right) \;,\;\;-\pi /R<q<\pi /R 
\label{Chi0(q_omega)1DAnsatz} \\
&\approx &\frac{BR^{2}}{2}\frac{q^{2}}{\omega ^{2}}\;\;\text{for }\left|
q\right| <<R^{-1},\;\;R=1D\;lattice\;spacing  \label{Chi0(lowq)1DAnsatz}
\end{eqnarray}%
This $q^{2}/\omega ^{2}$ dependence is the correct long-wavelength,
high-frequency form of the metallic density response from \ Bloch band
theory. \ \ Comparing (\ref{Chi0(lowq)1DAnsatz}) and (\ref{Chi0_1Dmetal}) we
find we can make our Ansatz give the correct long-wavelength response
provided that we choose the constant $B$ in $\chi _{0}^{Ansatz}$ such that 
\begin{equation}
\chi _{0}^{1D}\left( q,\omega \right) =\chi_0 ^{Ansatz}\left( q,\omega \right)
=\frac{1}{R}\chi_0 ^{Ansatz}\left[ q,\omega \right] \;\;\;for\;\;q\rightarrow 0
\label{CostraintOnChi0Ansatz}
\end{equation}
The factor of $1/R$ on the right side of (\ref{CostraintOnChi0Ansatz})
comes from the conversion between continuous '' ( )'' and discrete ''[ ]''
Fourier transforms.
 Putting (\ref{Chi0_1Dmetal}) and (\ref{Chi0(lowq)1DAnsatz}) into (\ref%
{CostraintOnChi0Ansatz}) gives 
\begin{equation}
\frac{N_sN_vv_{F}}{\pi \hbar }\frac{q^{2}}{\omega ^{2}}=\frac{1}{R}\frac{BR^{2}}{2}
\frac{q^{2}}{\omega ^{2}}\;\therefore \;B=\frac{2N_s N_v v_{F}}{\pi \hbar R}
\label{Bfor1DCase}
\end{equation}%
so (\ref{ChoAnsatz_qOmega_1D}) becomes%
\begin{equation}
\omega ^{2}\chi _{0}^{Ansatz,1D}[q,\omega ]=c_{0}\left( q\right) =\frac{%
2N_s N_v v_{F}}{\pi \hbar R}c_{00}\left( q\right) ,\;\;\;\;c_{00}\left( q\right)
=1-\cos \left( qR\right) =2\sin ^{2}\left( \frac{qR}{2}\right)
\label{ChoAnsatz_qOmega_1D}
\end{equation}%

This agrees with the known 1D metallic response when $q\rightarrow 0$ but is
not necessarily correct for larger $q$. \ However the principal ''Type-C''
effects of metallicity on dispersion interactions come from the
long-wavelength metallic response and so the present model should be
appropriate. \ To make a better Ansatz for the discrete response one
could include more distant neighbors in (\ref{GeneralNN_Chi0Ansatz}).\ The
coefficients of the next-neighbor terms could be chosen to obtain the
correct $\chi _{0}( q,\omega )$ through $O\left( q^{4}\right) ,$ for example.
This will not be done here, \ however.

\section{An atomic chain with both polarizable atoms and a metallic Bloch
band}

\ Here we use a 1D chain of atoms to demonstrate explicitly how the
discrete-atom Many Body Dispersion (MBD) approach 
 \cite{mbd,mbdrs,proof,Kim_uMBD_2020,science,jpcl,ts} can be generalized
to include a metallic band in the electronic structure. We will obtain an
analytic solution for the coupled modes of oscillation.

We consider a chain of ''atoms'' with uniform spacing $R$\ \ so that the
atoms have locations 
\[x_{J}=JR,\;\;J=0,\pm 1,\pm 2... 
\]%
On each site $J$ we assume there is an ''atom'' that is linearly polarizable
so that its transient electric dipole moment under the action of a small
electric field ${\cal \vec{E}}_{I}$ is%
\begin{equation}
\vec{p}_{I}e^{-i\omega t}={\bf \alpha }\left( \omega \right) {\cal \vec{E}}%
_{I}e^{-i\omega t},\;\;{\bf \alpha =}\alpha _{xx}\hat{x}\hat{x}+
\alpha _{yy}\hat{y}\hat{y}%
+\alpha _{zz}\hat{z}\hat{z}  \label{PiEqAlphaEi}
\end{equation}%
\ 
We have assumed that the principal axes of the atomic polarizability 
tensor $\bf{\alpha}$ lie along the $x$ direction pointing along the chain, 
and in two mutually perpendicular directions $y$ and $z$ perpendicular
 to the chain.\ The
electric potential and field at atom $J$ on the chain due to a point dipole 
$\vec{p}$ on site $I$ are 
\begin{eqnarray*}
\phi _{I} &=&\frac{\vec{r}\cdot \vec{p}}{R^{3}} \\
{\cal \vec{E}}_{{}} &{\cal =}&{\cal -}\frac{\vec{p}-3\left( \vec{p}\cdot \hat{r}%
\right) \hat{r}}{r^{3}},\;\;\vec{r}=\left( I-J\right) R\hat{x}
\end{eqnarray*}%
This shows that, on a single chain, the transverse polarizations do not
Coulomb-couple to each other \ or to the longitudinal polarizations.

In addition there is a metallic electron band so that the transient
excess electron number on each site is

\begin{equation}
n_{I}e^{-i\omega t}=\sum_{J}\chi _{0}[I-J,\omega ]v_{J}e^{-i\omega t}
\label{NiEqChiVi}
\end{equation}%
where $v_{J}$ is the electronic potential energy perturbation at site $J$,
and $\chi _{0}$ is given by (\ref{Chi0Ansatz)_1Dchain}).

We seek self-sustaining oscillations where there are no external driving
fields. \ We will treat the collective oscillations in the direct Random Phase
Approximation (dRPA), which starts from the independent-electron response 
$\chi_0$, then screens it within time-dependent mean field theory. \
In fact individual electron excitations contained in $\chi_{0}$ are
suppressed in 1D systems because of \ strong electron-electron Coulomb
effects. Nevertheless the long-wavelength collective excitations (1D
plasmons) are known \cite{DasSarma92} to be well described by the dRPA. It
is these plasmons that drive the Type-C dispersion energy effects.

\ In dRPA the internal fields (assuming point charges and dipoles on the
atoms) are%
\begin{equation}
v_{I}=\sum_{J}w_{I-J}n_{J}+u_{I-J}p_{J_{x}},\;\;\;\;{\cal E}%
_{Ix}=\sum_{J}u_{I-J}n_{J}+\sum_{J}T_{I-J}p_{Jx}
\label{RealSpaceFieldsOneChain}
\end{equation}%
\[
{\cal E}_{Iy}=\sum_{J}T_{I-J}^{\perp }p_{Jy},\;\;\;{\cal E}%
_{Iz}=\sum_{J}T_{I-J}^{\perp }p_{Jz}, 
\]%
Here the charge-charge, charge-dipole and dipole-dipole Coulomb interactions
are 
\begin{eqnarray}
w_{I-J} &=&w_{0}\delta _{I-J}+\frac{e^{2}}{R\left| I-J\right| }\left(
1-\delta _{IJ}\right) ,\;\;u_{I-J}=\frac{-\left| e\right| }{R^{2}}\frac{%
sgn\left( I-J\right) }{\left( I-J\right) ^{2}}\left( 1-\delta _{IJ}\right)
,\;\;  \label{ReaSpace1Dpotentials} \\
T_{I-J} &=&\frac{2}{R^{3}\left| I-J\right| ^{3}}\left( 1-\delta _{IJ}\right)
,\;\;T_{I-J}^{\perp }=-\frac{1}{2}T_{I-J}\;.
\end{eqnarray}

This point-dipole assumption is reasonable for the fields due to well
separated atoms but, as with the usual MBD theory, the nearest-neighbor
interactions will need to be modified. In (\ref{ReaSpace1Dpotentials}), $%
w_{0}$ is the self-interaction of the localized orbital that describes the
conduction band (e.g. a $\pi _{z}$ orbital for graphenics). \ We can specify 
$w_{0}$ by the parameter $R_{0}$:
\begin{equation}
w_{0}=w_{I=J}=\frac{e^{2}}{R_{0}}  \label{DefnR0}
\end{equation}%
Here $R_{0}$ characterizes the spatial extent of the atomic conduction
orbital. \ The self-interaction term $w_{0}$ contains a presumably unphysical
interaction of an electron with itself. This unphysical orbital
self-interaction is naturally present in the dRPA and is partly responsible
for the known tendency of dRPA to over-screen. However, in a non-magnetized band metal, any transient charge that appears on an atom is
composed equally of spin-up and spin-down electron density, and there is a
genuine Coulomb interaction energy between the two different spin species.
\ The unphysical part in dRPA is the on-site interaction of a given spin
species with itself.\ \ Thus $w_{0}$ and hence $R_{0}$ could\ in principle
be determined by evaluating Coulomb self-energy of a localized orbital from a
tight-binding description of the conducting band. \ If the unphysical same-spin
contribution is removed by halving this interaction, one has the potential to
describe the 1D plasmon better than the dRPA does.\ Indeed another way to
determine $w_{0}$ would be to fit the 1D plasmon dispersion at a large
wavenumber from a high-level microscopic calculation of the 1D plasmon
dispersion. \ In the present preliminary work, however$,R_{0}$, and hence 
$w_{0}$ will be treated as an adjustable parameter.

Regardless \ of how its numerical value is specified, the on-site interaction 
$w_{0}$ from Eq. (\ref{DefnR0}) is needed, in the present discrete approach,
to prevent an instability whereby the one-dimensional plasmon frequency
becomes imaginary at larger wavenumbers, \ as will be shown explicitly below.

Now we apply discrete Fourier-transformation to the dRPA equations(\ref%
{PiEqAlphaEi}). There is a convolution theorem\ for this type of Fourier
transform  and the relations
 (\ref{NiEqChiVi}), (\ref{PiEqAlphaEi}) and (\ref{RealSpaceFieldsOneChain}) become 
\begin{eqnarray}
n\left( q\right) &=&\chi _{0}^{Ansatz}\left( q\right) \left( w\left(
q\right) n\left( q\right) +u\left( q\right) p_{x}\left( q\right) \right) 
\nonumber \\
p_{x}\left( q\right) &=&\alpha _{xx}\left( \omega \right) \left( u\left(
q\right) n\left( q)+T\left( q\right) p_{x}\left( q\right) \right) \right) \\
p_{y}\left( q\right) &=&\alpha _{yy}\left( \omega \right) T^{\perp }\left(
q\right) p_{y}\left( q\right) \\
p_{z}\left( q\right) &=&\alpha _{zz}\left( \omega \right) T^{\perp }\left(
q\right) p_{z}\left( q\right)  \label{RPAEquations1Chain}
\end{eqnarray}

Here the transformed interactions are 
\begin{equation}
w\left( q\right) =\frac{e^{2}}{R}w_{00}\left( q\right)
,\;\;\;\;\;w_{00}\left( q\right) \equiv \frac{R}{R_{0}}+\sum_{m=1}^{\infty }%
\frac{2\cos \left( qRm\right) }{m}=\frac{R}{R_{0}}-\ln \left( 2\left( 1-\cos
\left( qR\right) \right) \right) \;\;  \label{Defn_w00}
\end{equation}%
(see Gradsteyn and Ryzhik \cite{Gradshteyn_Ryzhik} formula $1.441(2)).$%
\begin{equation}
u\left( q\right) =i\frac{\left| e\right| }{R^{2}}u_{00}\left( q\right)
,\;\;\;\;u_{00}\left( q\right) \equiv \sum_{m=1}^{\infty }\frac{2\sin \left(
qRm\right) }{m^{2}}=2\int_{0}^{qR}\left( -\frac{1}{2}\ln \left( 2\left(
1-\cos x\right) \right) \right) dx  \label{Defn_u00}
\end{equation}%
\begin{eqnarray}
T\left( q\right) &=&\frac{1}{R^{3}}T_{00}\left( q\right) ,\;\;\;T_{00}\left(
q\right) \equiv \sum_{m=1}^{\infty }\frac{4\cos \left( qRm\right) }{m^{3}}%
,\;\;\;\;  \label{Defb_T00} \\
T^{\perp }\left( q\right) &=&\frac{1}{R^{3}}T_{00}^{\perp }\left( q\right)
,\;\;\;T_{00}^{\perp }\left( q\right) =-\frac{1}{2}T_{00}\left( q\right)
\end{eqnarray}%
and $\chi _{0}^{Ansatz}\left( q\right) $ is given by (\ref%
{Chi0(q_omega)1DAnsatz}). \ \ \ 

The dimensionless interactions $T_{00},\;T_{00}^{\perp }$ and $u_{00}$ can
be evaluated numerically and have the following properties. \ 

\begin{eqnarray}
T_{00}\left( q_{1}\right) &<&T_{00}\left( q_{2}\right)
\;\;when\;\;q_{1}>q_{2}>0  \nonumber \\
\,T_{00}\left( 0\right) &=&4.\,\allowbreak 808\,227\,6,\;\;T_{00}\left( \pi
/R\right) =-3.\,\allowbreak 606\,170\,7  \label{T00Values}
\end{eqnarray}%
\begin{eqnarray}
T_{00}^{\perp }\left( q_{1}\right) &>&T_{00}^{\perp }\left( q_{2}\right)
\;\;when\;\;q_{1}>q_{2}>0  \nonumber \\
\,T_{00}^{\perp }\left( 0\right) &=&-2.\,\allowbreak
404\,112\,8,\;\;T_{00}^{\perp }\left( \pi /R\right) =1.\,\allowbreak
803\,085\,4  \label{T00PerpVlues}
\end{eqnarray}%
\begin{eqnarray}
u_{00}\left( 0\right) &=&u_{00}\left( \pi /R\right) =0  \nonumber \\
0 &\leq &u_{00}\left( q\right) \lessapprox u_{00}\left( 1.2/R\right) \approx
2.02  \label{u00Values}
\end{eqnarray}

We incorporate a simple version of MBD by assuming a single 3D harmonic
oscillator on each atom, all with the same resonant frequencies $\omega
_{xx,} $ $\omega _{yy},\omega _{zz}$ and then we can write

\[
\alpha _{ii}\left( \omega \right) =\alpha _{0ii}\frac{\omega _{ii}^{2}}{%
\omega _{ii}^{2}-\omega ^{2}},\;\;\chi _{0}^{Ansatz}[q,\omega ]=\frac{%
c_{0}\left( q\right) }{\omega ^{2}},\;\;\;c_{0}\left( q\right) =\frac{4N_s N_v v_{F}%
}{\pi \hbar R}\sin ^{2}\left( \frac{qR}{2}\right) 
\]%

where $a_{0ii}$ is the static polarizability of an ''atom'' in the $i^{th}$ %
spatial direction.

The last two equations of Eqs. $\left( \text{\ref{RPAEquations1Chain}}%
\right)$ do not couple, and the first two can then be written%
\begin{equation}
\left( 
\begin{array}{cc}
\frac{c_{0}}{\omega ^{2}}w-1 & \frac{c_{0}}{\omega ^{2}}u \\ 
\frac{\alpha _{xx}\omega _{xx}^{2}}{\omega _{xx}^{2}-\omega ^{2}}u & \frac{%
\alpha _{xx}\omega _{xx}^{2}}{\omega_{xx}^{2}-\omega ^{2}}T-1
\end{array}%
\right) \left( 
\begin{array}{c}
n \\ 
p_{x}%
\end{array}%
\right) =\left( 
\begin{array}{c}
0 \\ 
0%
\end{array}%
\right)  \label{2By2_matrix_1Chain2}
\end{equation}%

while the last two give
\begin{equation}
\left( \frac{\alpha _{yy}\omega _{yy}^{2}}{\omega _{yy}^{2}-\omega ^{2}}%
T^{\perp }-1\right) p_{yy}=0,\;\;\;\left( \frac{\alpha _{zz}\omega
_{zz}^{2}}{\omega _{zz}^{2}-\omega ^{2}}T^{\perp }-1\right) p_{zz}=0
\label{Transverse_ModeEqs_1Chain}
\end{equation}

\subsection{Decoupled charge and polarization modes of a single chain}

If we switch off the charge-dipole coupling term - i.e. set $u=0$ $\ $in (\ref
{2By2_matrix_1Chain2}) - we find four separate charge and dipolar modes $
\omega_{10,},\omega _{20}$,$\;\omega _{30},\;\omega _{40}:$

The first mode is a plasma wave involving the movement of charge between
atoms:%
\begin{eqnarray}
\omega _{10}^{2}\left( q\right) &=&c_{0}\left( q\right) w\left( q\right)
=\left( \frac{4N_s N_vv_{F}}{\pi \hbar R}\sin ^{2}\left( \frac{qR}{2}\right)
\right) \frac{e^{2}}{R}\left( -\ln \left( 2\left( 1-\cos \left( qR\right)
\right) \right) +\frac{R}{R_{0}}\right)  \nonumber \\
&=&\left( \frac{4N_s N_v v_{F}e^{2}}{\pi \hbar R^{2}}\sin ^{2}\left( \frac{qR}{2}%
\right) \right) \left( -\ln \left( 2\left( 1-\cos \left( qR\right) \right)
\right) +\frac{R}{R_{0}}\right)  \nonumber \\
&=&\frac{4N_s N_v v_{F}e^{2}}{\pi \hbar R^{2}}\sin ^{2}\left( \frac{qR}{2}\right) %
\left[ -\ln \left( 4\sin ^{2}\left( \frac{qR}{2}\right) \right) +\frac{R}{%
R_{0}}\right]  \label{1DPlasmonArbq} \\
\omega _{10}\left( q\right) &\approx &\sqrt{\frac{N_s N_v v_{F}e^{2}}{\pi \hbar }}q%
\sqrt{-2\ln \left( qR\right) }\;\;\;for\;\;qR\rightarrow 0\;\;\text{where\ \
the logarithm dominates.}  \nonumber
\end{eqnarray}%

This quasi-acoustic ($\omega_{10}\propto q\sqrt{|\ln q|})\;$mode
is the well-known one-dimensional plasmon arising from the conduction band.\
\

The last three of these uncoupled modes are polarization waves:

\begin{equation}
\omega _{20}\left( q\right) =\sqrt{\omega _{xx}^{2}-a_{xx}\omega
_{xx}^{2}T\left( q\right) }=\omega _{xx}\sqrt{1-\alpha
_{xx}T_{00}\left( q\right) }  \label{LongitudPolarizationModeFrequ}
\end{equation}%
\begin{equation}
\omega _{30}\left( q\right) =\omega _{yy}\sqrt{1-\alpha _{yy}T^{\perp
}\left( q\right) },\;\;\omega _{40}\left( q\right) =\omega _{zz}\sqrt{%
1-\alpha _{zz}T^{\perp }\left( q\right) }
\label{TransversePolariztionModeFrequencies}
\end{equation}%
These three are the MBD mode describing polarization waves arising from
coupled localized atomic oscillators. \ These modes have the well-known
problem that the frequency can become imaginary if $a_{0}$ is too large (the
''polarization catastrophe'', suggesting a ferroelectric instability). This
arises partly because the formalism has assumed point dipoles. \ That problem  is
typically remedied by modifying the Coulomb interaction at short distance,
using parameters optimized by fitting dispersion energies from a
dispersion-oriented benchmark data set. \ This has the advantage of
indirectly accommodating phenomena such as bond formation, which are not
directly envisioned in the MBD approach.

It is convenient to express all mode frequencies in terms of the MBD atomic
resonant frequencies

\begin{eqnarray}
\omega _{10} &=&\omega _{xx}\sqrt{S_{1}\left( 1-\cos \left( qR\right)
\right) \left[ -\ln \left( 4\sin ^{2}\left( \frac{qR}{2}\right) \right) +%
\frac{R}{R_{0}}\right]}  \nonumber \\
\omega _{20} &=&\omega _{xx}\sqrt{1-S_{2}T_{00}\left( q\right) }  \nonumber
\\
\omega _{30} &=&\omega _{yy}\sqrt{1-S_{3}T_{00}^{\perp }\left( q\right) } 
\nonumber \\
\omega _{40} &=&\omega _{zz}\sqrt{1-S_{4}T_{00}^{\perp }\left( q\right) }
\label{DecoupledModeFreqs_dimless}
\end{eqnarray}
where $T_{00}$ and $T_{00}^{\perp }$ are given by (\ref{Defb_T00}) and the dimensionless constants $S$ are defined by%
\begin{equation}
S_{1}=\frac{2Ns N_v v_{F}e^{2}}{\pi \hbar R^{2}\omega _{xx}^{2}}\;\;\;S_{2}=\frac{%
\alpha _{xx}}{R^{3}},\;\ S_{3}=\frac{\alpha _{yy}}{R^{3}}\ \;\;S_{4}=\frac{%
\alpha _{zz}}{R^{3}}.\;\;\;\;\;\;\;\;\;\;  \label{Defns_S1_S2_S3_S4}
\end{equation}

\subsection{Stability of the decoupled polarization and plasmon modes of a
single chain}

If the point-source form of the Coulomb interactions is kept unmodified, the
conditions for stable polarization modes (i.e. real mode  frequencies, $%
\omega ^{2}\geq 0$ in Eqs (\ref{LongitudPolarizationModeFrequ},\ref%
{TransversePolariztionModeFrequencies})) are as follows (see also Eqns. (\ref%
{T00Values},\ref{T00PerpVlues})):%
\begin{equation}
\;\;\alpha _{xx}<R^{3}/\max_{q}T_{00}\left( q\right) =R^{3}/T_{00}\left(
q=0\right) =0.207\,976\,84R^{3}\;\;\Rightarrow \;\;S_{2}\leq
0.207\,976\,84\;\;  \label{StabilityCond_x}
\end{equation}%
\begin{equation}
\alpha _{yy},\alpha _{zz}<R^{3}/\max_{q}T_{00}^{\perp }\left( q\right)
=R^{3}/1.\,\allowbreak 803\,085=0.554\,6R^{3}\;\;\Rightarrow
\;S_{3},S_{4}\leq 0.554\,6\;\;  \label{StabilityCond_y_z}
\end{equation}

The 1D plasmon mode (\ref{1DPlasmonArbq}) is stable, even with point Coulomb
interactions for $I\neq J$, provided%
\begin{equation}
R/R_{0}>\ln 4=1.\,\allowbreak 386\,294\,4  \label{StabiityCond_plasmon}
\end{equation}

\begin{figure}[tbp]
\includegraphics[width=0.9\textwidth]{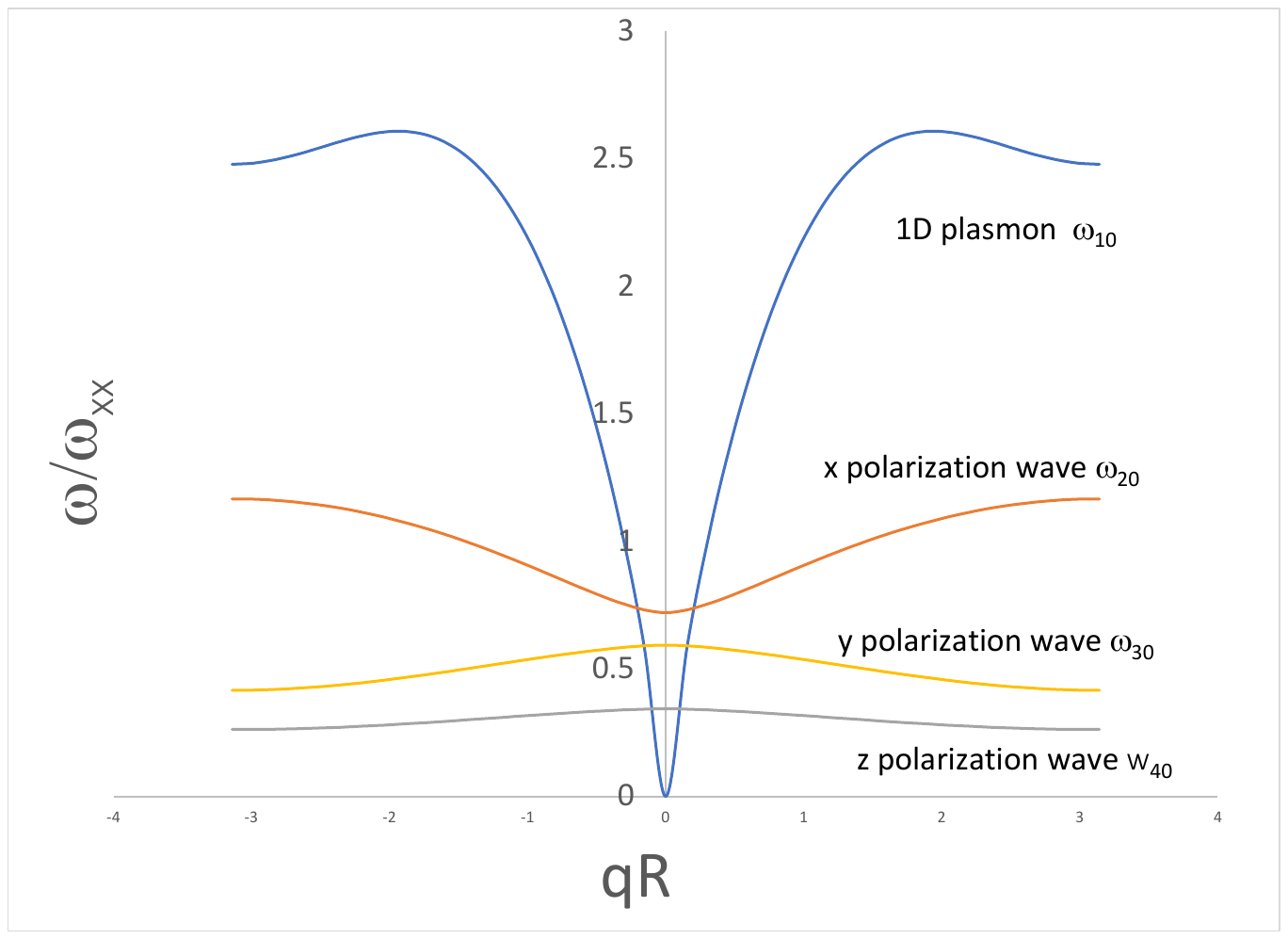}
\caption{
Dimensionless uncoupled mode frequencies $ \omega / \omega_{xx}$ for  a single chain that is  strongly metallic ($S_1  =5$ ) and moderately and anisotropically polarizable ($S_2 = 0.1,  S_3=0.13, S_4=0.17, \omega_{yy}/\omega_{xx}=0.3, \omega_{zz}/\omega_{xx}=0.5  $), with $R/R_0= 2$.  Here the Coulomb interaction $u$ between the charge mode (1D plasmon) and the x-polarization mode has been switched off.
\label{Fig1UncoupledFreqs} 
}
\end{figure}

The decoupled ($u=0$) mode frequencies $\omega _{j0},\;\;j=1,...4$ are plotted vs. 
dimensionless wavenumber $qR$ in Fig  \ref{Fig1UncoupledFreqs}.  Dimensionless 
 forms,  $\omega$ / $\omega_{xx}$, are shown, where input parameter  $\omega_{xx}$.
is the  oscillator frequency  of the atoms in the $x$ direction along the chain.
  The parameters were chosen to 
represent a chain that is strongly  metallic ($S_{1}=5$) and moderately
polarizable ($S_{2}=0.1,S_{3}=0.13, S_4 =0.17$).  For this case the plasmon and
longitudinal polarization modes cross, but this is not the case for more
weakly metallic chains. \ The transverse polarization modes are
qualitatively similar to the longitudinal one, but their frequency decreases
rather than increasing with increasing $\left| q\right| $.  These behaviors 
lead,  for large atomic polarizabilities, to mode instabilities ($\omega^2 < 0$) 
that occur first at the Brillouin zone center for the x-polarization mode, and 
near the zone boundary for the y- and z- modes.    The parameters $S_3=0.13$, 
 $S_4=0.17$,   $ \omega_{yy} $ / $\omega_{xx} =0.5$, 
$ \omega_{zz}$ / $\omega_{xx} = 0.3$ correspond to atoms with different polarizabilities
and resonant frequencies in response to fields in different spatial directions.  The need for this 
flexibility will become apparent in later Sections where we discuss chains of gold atoms.

\subsection{Coupled longitudinal mode frequencies of one chain}

When the charge-dipole Coulomb interaction $u\left( q\right) $ is nonzero,
it couples the 1D plasmon to the longitudinal polarization mode but leaves
the transverse (y- and z-directed) polarization modes unaffected.

We multiply the top line of\ (\ref{2By2_matrix_1Chain2}) by $\omega ^{2}$ and
the second line by $\left( \omega ^{2}-\omega _{0}^{2}\right) $. This gives 
\[
\left( 
\begin{array}{cc}
c_{0}w-\omega ^{2} & c_{0}u \\ 
-\alpha _{xx}\omega _{xx}^{2}u & -\alpha _{xx}\omega _{xx}^{2}T+\omega
_{xx}^{2}-\omega ^{2}%
\end{array}%
\right) \left( 
\begin{array}{c}
n \\ 
p%
\end{array}%
\right) =\left( 
\begin{array}{c}
0 \\ 
0%
\end{array}%
\right) 
\]

with nonzero solutions only if%
\[
0=\omega ^{4}-\left( c_{0}w-\alpha _{xx}\omega _{xx}^{2}T+\omega
_{0}^{2}\right) \omega ^{2}+c_{0}w\left( -\alpha _{xx}\omega _{xx}^{2}T+\omega
_{xx}^{2}\right) +c_{0}\alpha _{xx}\omega _{xx}^{2}u^{2} 
\]%
There are two longitudinal-mode frequencies $\omega _{1}\left( q\right) $, $%
\omega _{2}\left( q\right) $ for each $q$ :%
\[
\left\{ 
\begin{array}{c}
\omega _{1}^{2} \\ 
\omega _{2}^{2}%
\end{array}%
\right\} =\frac{1}{2}\left( 
\left( \omega _{xx}^{2}-\alpha _{xx}\omega _{xx}^{2}T+c_{0}w\right) \mp
\sqrt{\left( \omega _{xx}^{2}-\alpha _{xx}\omega _{xx}^{2}T-c_{0}w\right)
^{2}-4\alpha _{xx}\omega _{xx}^{2}c_{0}u^{2}} \right)%
\]

Using the real dimensionless interaction quantities $w_{00}\left( q\right)
,u_{00}\left( q\right) ,$\ and $T_{00}\left( q\right) $ defined in Eqs (\ref%
{Defn_w00})-(\ref{Defb_T00}) and \ the dimensionless susceptibility $%
c_{00}\left( q\right) $ defined in (\ref{ChoAnsatz_qOmega_1D}) we can write
the eigenfrequencies of the coupled charge-dipole modes of a single chain in
dimensionless form: 
\begin{equation}
\left\{ 
\begin{array}{c}
\omega _{1}^{2} \\ 
\omega _{2}^{2}%
\end{array}%
\right\} =\frac{\omega _{xx}^{2}}{2}\left( \left(
1+S_{1}c_{00}w_{00}-S_{2}T_{00}\right) \mp \sqrt{\left(
1-S_{1}c_{00}w_{00}-S_{2}T_{00}\right) ^{2}+4S_{2}S_{1}c_{00}u_{00}^{2}}%
\right)  \label{CoupledModes_1Chain}
\end{equation}%
Here the ''$_{00}$'' quantities are dimensionless functions of wavenumber $q$%
. The dimensionless parameters $S_{j}$ are given by (\ref{Defns_S1_S2_S3_S4}%
). Larger values of $S_{2}$,S$_{3},S_{4}$ imply more polarizable atoms. A
larger value of $S_{1}$ makes the metallic character more important
relative to the parallel atomic polarizability.

\subsection{Stability: are the 1-chain coupled mode frequencies real ?}

We will now show that the point-Coulomb-coupled modes (\ref%
{CoupledModes_1Chain}) are stable (i.e. they have real frequencies) provided
that conditions (\ref{StabiityCond_plasmon},\ref{StabilityCond_x}) are met
-i.e. that the  uncoupled plasmon and longitudinal polarization modes are
stable. The square root $\sqrt{b}$ on the right side of (\ref%
{CoupledModes_1Chain}) is always real because its argument $b$ is a sum of
squares. \ But is $\sqrt{b}\;$small enough to keep $\omega _{1}^{2}>0$? By
completing a square, we transform the square root argument to%
\begin{eqnarray}
b &=&\left( 1-S_{2}T_{00}-S_{1}c_{00}w_{00}\right)
^{2}+4S_{1}S_{2}c_{00}u_{00}^{2}\;\; \\
&=&A^{2}-4S_{1}c_{00}\left\{ w_{00}\left( 1-S_{2}T_{00}\right)
-S_{2}u_{00}^{2}\right\} \;  \label{SqrtArgument}
\end{eqnarray}%
where $A=1+S_{1}c_{00}w_{00}-S_{2}T_{00}$\ . \ $A\,$is positive provided
that the uncoupled stability conditions (\ref{StabiityCond_plasmon}) and (%
\ref{StabilityCond_x}) are satisfied. Under these same conditions the
expression $Y\left( q\right) $ in braces in (\ref{SqrtArgument}) satisfies

\[
\left\{ {}\right\} \equiv Y\left( a\right) >X\left( \text{q}\right) =\ln
\left( \frac{2}{1-\cos \left( qR\right) }\right) \left(
1-0.207\,976\,84T_{00}\left( q\right) \right)
-0.207\,976\,84u_{00}^{2}\left( q\right) 
\]%
The quantity $X\left( q\right) $ is independent of the system parameters $%
\alpha _{jj,\;}$, $\omega _{0ji}$ and $v_{F}$ - \ i.e. applicable to all
monoatomic chains, We evaluated $X\left( q\right) $ numerically and found it
to be non-negative in the entire Brillouin zone - specifically $0\leq X\left(
q\right) <6.9\times 10^{-2}$ for $-\pi /R\leq q\leq /\pi R.\;$Thus the
squared coupled longitudinal mode frequencies (\ref{CoupledModes_1Chain})
satisfy

\[
\left\{ 
\begin{array}{c}
\omega _{1}^{2} \\ 
\omega _{2}^{2}%
\end{array}%
\right\} =A\mp \sqrt{A^{2}-4S_{1} c_{00}Y\left( q\right) }\geq 0 
\]
where $4S_{1} c_{00}Y\left( q\right) \geq 0$
 because $S_{1}>0$, $c_{00}\geq0$ and $Y\left( q\right) \geq X(q)\geq 0.$

Thus provided that the uncoupled longitudinal modes of a single chain are
stable, the coupled modes are stable also.

\begin{figure}[tbp]

\includegraphics[width=0.9\textwidth]{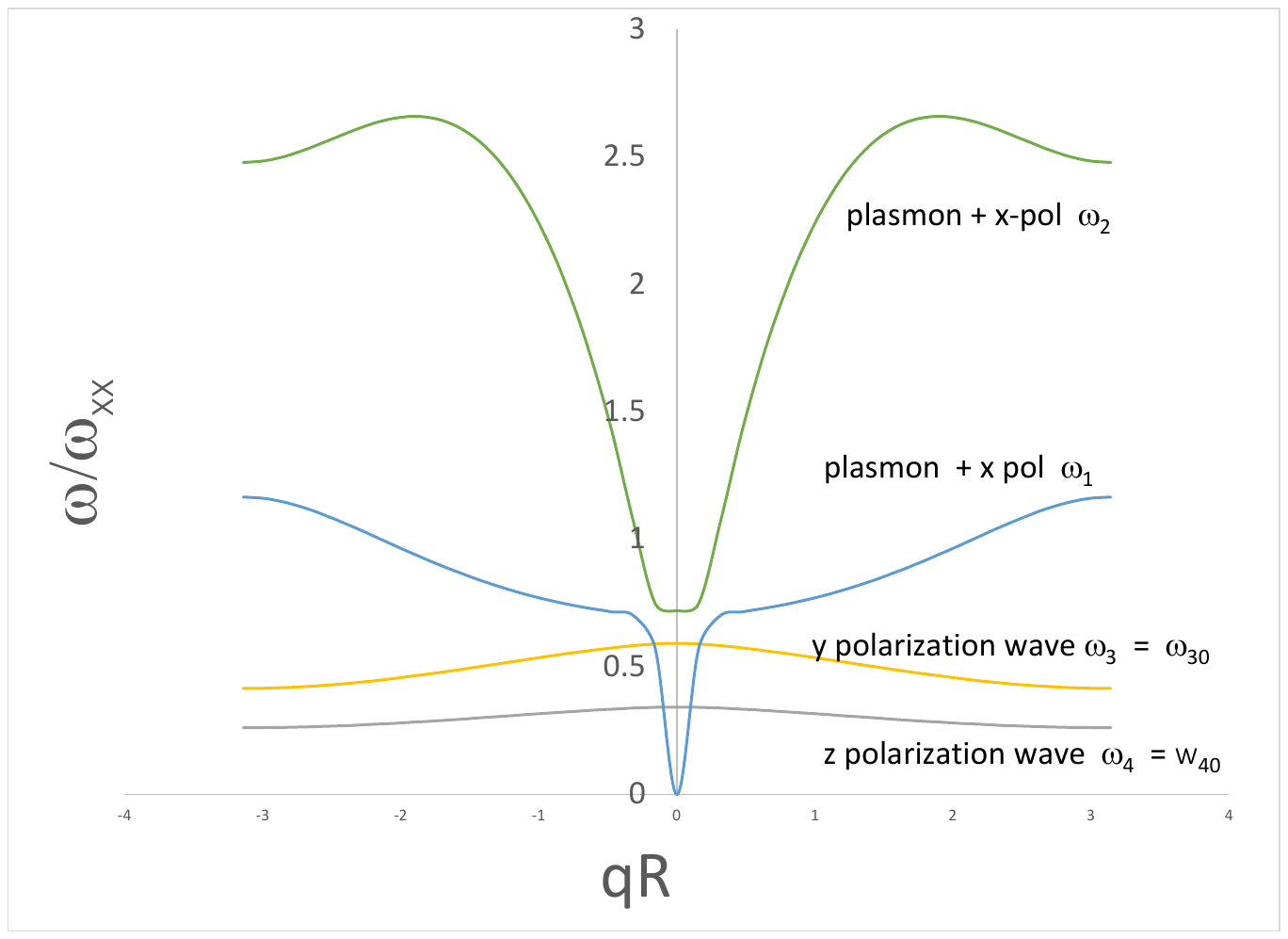}
\caption{
Dimensionless mode frequencies $ \omega / \omega_{xx}$ for a single chain that is 
strongly metallic and moderately polarizable, with the same parameters as in Fig
 \ref{Fig1UncoupledFreqs}. 
  In contrast to Fig. \ref{Fig1UncoupledFreqs},  however, here  the Coulomb  coupling
$u$ between charges and x-dipoles has been switched on.  Note the avoided crossing between
 plasmon and x-dipole modes, which nevertheless do not strongly modify each other away from
 the crossing point.  The y-and z-dipole modes do not  Coulomb-couple to other modes, 
and are the same  as in Fig. 1.  All modes are stable with the chosen parameter values.
\label{CpledModes1Chain}
}
\end{figure}

The coupled mode frequencies of a single chain are plotted in Fig. 
\ref{CpledModes1Chain} for the 
same parameters as in Fig \ref{Fig1UncoupledFreqs}
  except that the charge-dipole Coulomb  interaction 
$u$ is now included.
This chain  is  strongly metallic, but is still within the
region of mode  stability. \ Note the avoided crossing. Apart from the region near
the crossing, the charge and polarization modes do not affect each other very
much, at least for the chosen values of the parameters $S_{1},\;S_{2}$. 
 The y- and z-polarization modes on a single chain are the same regardless of the value of the 
charge-polarization Coulomb coupling $u$, and thus are the same as in Figure 
\ref{Fig1UncoupledFreqs}.

The fact that the present theory obtains the quasi-acoustic 1D plasmon is
essential for the correct inclusion of Type-C nonadditivity and unusual
power laws for the spatial decay of the dispersion interaction between chains, as
presented below.

\section{Two parallel chains of atoms: analytic mode frequencies for
parallel-only atomic polarizabilities.}

Consider two infinite parallel chains of atoms separated by distance $D,$
with uniform spacing $R$ between atoms within each chain. \ For definiteness
we treat the case of ''on-top registry'' where the positions $%
X_{I}^{(1)},\;X_{I}^{\left( 2\right) }$ of atoms on the two chains are 
\[
\vec{X}_{I}^{\left( 1\right) }=IR\hat{x},\;\;X_{I}^{\left( 2\right) }=IR\hat{%
x}+D\hat{y} 
\]%
As for the case of 1 chain, we allow each atom to have a transient charge $%
q_{I}^{\left( m\right) }\exp \left( -i\omega t\right) \equiv -\left|
e\right| n_{I}^{\left( m\right) }\exp \left( -i\omega t\right) $ and a
transient dipole $p_{I}^{\left( m\right) }\exp \left( -i\omega t\right) $
directed along the chain. \ Here $m=1$ or $2$ labels the two chains. \ The
interactions $w\left[ I-J\right] $, u$\left[ I-J\right] $ and T$\left[ I-J%
\right] $ among charges and dipoles within each chain are the same as
described in the previous paragraphs. \ In the present Section we
assume that the atomic polarizability is nonzero only in the  ''parallel" 
direction (along the chain): this restriction leads to an analytic solution for
the modes. Then the interactions between atoms $I$ on one chain and $J$ on
the other chain are, \ with\ $j\equiv I-J=0,\pm 1,\pm 2,..$ and $d\equiv D/R$%
\begin{equation}
w(D,I-J]=\frac{e^{2}}{\left( j^{2}R^{2}+D^{2}\right) ^{1/2},}%
\;,\;\;w\left( D,q\right) =\frac{e^{2}}{R}w_{0D}\left( q,\right)
,\;\;w_{0D}\left( q\right) =\frac{1}{d}+\sum_{j=1}^{\infty }\frac{2\cos
\left( jqR\right) }{\left( j^{2}+d^{2}\right) ^{1/2}}  \label{Defn_w0D}
\end{equation}

\begin{equation}
u(D,I-J]=\frac{-\left| e\right| jR}{\left( j^{2}R^{2}+D^{2}\right) ^{3/2}}%
\;,\;\;\;u(D,q)=i\frac{\left| e\right| }{R^{2}}u_{0D}\left( q\right)
,\;\;\;u_{0D}\left( q\right) =\sum_{j=1}^{\infty }\frac{2j\sin \left(
jqR\right) }{\left( j^{2}+d^{2}\right) ^{3/2}}\;\;\;  \label{Defn_u0D}
\end{equation}%
\begin{eqnarray}
T(D,I-J] &=&\frac{2j^{2}R^{2}-D^{2}}{\left( j^{2}R^{2}+D^{2}\right) ^{5/2}}
\;\;,\;\;\;T\left( D,q\right) =\frac{1}{R^{3}}T_{0D}\left( q\right) ,\;
\label{Defn_T0D} \\
\;\;T_{0D}\left( q\right) &=&\frac{-1}{d^{3}}+2\sum_{j=1}^{\infty }\frac{%
\left( 2j^{2}-d^{2}\right) \cos \left( jqR\right) }{\left(
j^{2}+d^{2}\right) ^{5/2}}
\end{eqnarray}

The sum in (\ref{Defn_w0D}), while finite, is poorly convergent so we
re-express it in terms of the known result for $D=0$ from (\ref{Defn_w00}):

\begin{equation}
w_{0D}\left( q\right) =\frac{1}{d}-\ln{\left(2 \left(1-\cos \left( qR\right)\right)
\right)} +\sum_{j=1}^{\infty }2\cos \left( jqR\right) \frac{1-\left( 1+\frac{%
d^{2}}{j^{2}}\right) ^{1/2}}{\left( j^{2}+d^{2}\right) ^{1/2}}
\label{W0D_Resummed}
\end{equation}

where the sum now converges like $\sum_{m}$ cos$\left( qRm\right) m^{-3}$.

In discrete Fourier transform space the RPA equations for charges and
dipoles on the two chains are%
\begin{eqnarray}
n^{\left( 1\right) }\left( q\right) &=&\frac{c_{0}\left( q\right) }{\omega
^{2}}\left( w\left( q,0\right) n^{\left( 1\right) }\left( q\right) +w\left(
q,D\right) n^{\left( 2\right) }\left( q\right) +u\left( q,0\right)
p^{\left( 1\right) }\left( q\right) +u\left( q,D\right) p^{\left( 2\right)
}\left( q\right) \right)  \nonumber \\
p^{\left( 1\right) }\left( q\right) &=&\frac{\alpha _{xx}\omega _{xx}^{2}}{%
\omega _{xx}^{2}-\omega ^{2}}\left( u\left( q,0\right) n^{\left( 1\right)
}\left( q\right) +u\left( q,D\right) n^{\left( 2\right) }\left( q\right)
+T\left( q,0\right) p^{\left( 1\right) }\left( q\right) +T\left( q,D\right)
p^{\left( 2\right) }\left( q\right) \right)  \nonumber \\
n^{\left( 2\right) }\left( q\right) &=&\frac{c_{0}\left( q\right) }{\omega
^{2}}\left( w\left( q,0\right) n^{\left( 2\right) }\left( q\right) +w\left(
q,D\right) n^{\left( 1\right) }\left( q\right) +u\left( q,0\right) p^{\left(
2\right) }\left( q\right) +u\left( q,D\right) p^{\left( 1\right) }\left(
q\right) \right)  \nonumber \\
p^{\left( 2\right) }\left( q\right) &=&\frac{\alpha _{xx}\omega _{xx}^{2}}{%
\omega _{xx}^{2}-\omega ^{2}}\left( u\left( q,0\right) n^{\left( 2\right)
}\left( q\right) +u\left( q,D\right) n^{\left( 1\right) }\left( q\right)
+T\left( q,0\right) p^{\left( 2\right) }\left( q\right) +T\left( q,D\right)
p^{\left( 1\right) }\left( q\right) \right)  \label{RPA_Eqs_2Wires_1st}
\end{eqnarray}

Define new variables%
\[
\eta ^{\left( 1,2\right) }\left( q\right) =c_{0}\left( q\right)
^{-1/2}n^{\left( 1,2\right) }\left( q\right) ,\;\;\pi ^{\left( 1,2\right)
}\left( q\right) =\alpha _{xx}^{-1/2}\omega _{xx}^{-1}p^{\left( 1,2\right)
}\left( q\right) 
\]%
and multiply Equations (\ref{RPA_Eqs_2Wires_1st}) through by $\omega
^{2},\;\omega _{xx}^{2}-\omega ^{2}$, $\omega ^{2},\;\omega _{xx}^{2}-\omega
^{2}$ respectively. The (\ref{RPA_Eqs_2Wires_1st}) becomes%
\[
\omega ^{2}\left( 
\begin{array}{c}
\rho ^{\left( 1\right) } \\ 
\pi ^{\left( 1\right) } \\ 
\rho ^{\left( 2\right) } \\ 
\pi ^{\left( 2\right) }%
\end{array}%
\right) =\omega _{xx}^{2}{\bf M}\left( 
\begin{array}{c}
\rho ^{\left( 1\right) } \\ 
\pi ^{\left( 1\right) } \\ 
\rho ^{\left( 2\right) } \\ 
\pi ^{\left( 2\right) }%
\end{array}%
\right) 
\]%
where ${\bf M}$ is hermitian and dimensionless:

\[
{\bf M}=\left( 
\begin{array}{cccc}
F & i\left| G\right| & f & i\left| g\right| \\ 
-i\left| G\right| & H & -i\left| g\right| & h \\ 
f & i\left| g\right| & F & i\left| G\right| \\ 
-i\left| g\right| & h & -i\left| G\right| & H%
\end{array}%
\right) \;\;. 
\]%
Here%
\begin{equation}
F=\frac{c_{0}w}{\omega _{xx}^{2}}=S_{1}c_{00}w_{00}\;\;,\;\;\;\;G=\frac{%
c_{0}^{1/2}\alpha_{xx}^{1/2}u}{i\omega _{xx}}=\sqrt{S_{1}S_{2}}c_{00}^{1/2}u_{00}\;%
\;\;,\;\;\;\;H=1-\alpha _{xx}T=1-S_{2}T_{00},\;\;\;\;\;
\end{equation}%
\begin{equation}
f=\frac{c_{0}w_{D}}{\omega _{xx}^{2}}=S_{1}c_{00}w_{0D}\;\;\;\;\;g=\frac{%
c_{0}^{1/2}\alpha_{xx}^{1/2}u_{D}}{i\omega _{xx}}=\sqrt{S_{1}S_{2}}%
c_{00}^{1/2}u_{0D},\;\;\;\;h\;=-\alpha _{xx0}T_{D}=-S_{2}T_{0D}
\label{Defn_f_g_h}
\end{equation}%
where $S_{1}$ and $S_{2}$ are given by (\ref{Defns_S1_S2_S3_S4}) and the other
quantities are defined by (\ref{ChoAnsatz_qOmega_1D}), (\ref{Defn_w00}) -(%
\ref{Defb_T00}), \ (\ref{Defn_w0D}) - (\ref{W0D_Resummed}). \ Further define 
\begin{eqnarray*}
\alpha &=&F-f,\;\;\;\;\;\gamma =H-h,\;\;\;\;\;\mu =F+f,\;\;\;\;\;\nu
=H+h,\;\;\; \\
\beta &=&\left| G\right| -\left| g\right| ,\;\;\;\phi =\left| G\right|
+\left| g\right| \;\;\;\;r=\sqrt{\left( \alpha -\gamma \right) ^{2}+4\beta
^{2}},\;\;\;\;s=\sqrt{\left( \mu -\nu \right) ^{2}+4\phi ^{2}}
\end{eqnarray*}%
where the  + and - signs in $\alpha$ ... $\phi$ correspond to even and odd modes as required by the mirror symmetry of  the  parallel-chain geometry.   
The four eigenvalues of ${\bf M}$ then give the squares of the coupled mode
frequencies of the pair of chains:%
\begin{eqnarray}
\frac{\omega _{1}^{2}}{\omega _{0}^{2}} &\equiv &\lambda _{1}=\frac{1}{2}%
\left( \alpha +\gamma +r\right) ,\;\;\;\;\;\;\frac{\omega _{2}^{2}}{\omega
_{0}^{2}}\equiv \lambda _{2}=\frac{1}{2}\left( \alpha +\gamma -r\right) \\
\frac{\omega _{3}^{2}}{\omega _{0}^{2}} &\equiv &\lambda _{3}=\frac{1}{2}%
\left( \mu +\nu +s\right) ,\;\;\;\;\;\;\frac{\omega _{4}^{2}}{\omega _{0}^{2}%
}\equiv \lambda _{4}=\frac{1}{2}\left( \mu +\nu -s\right)
\label{Coupled_2_chain_frequs}
\end{eqnarray}%
The four mod frequencies $\omega _{i}$ are plotted vs. dimensionless
wavenumber $qR$ in Fig. 3 for the case $d\equiv D/R=3.0,\;\;S_{1}=S_{2}=0.2$,  
$R/R_{0}\;=2$.

\begin{figure}[tbp]
\includegraphics[width=0.9\textwidth]{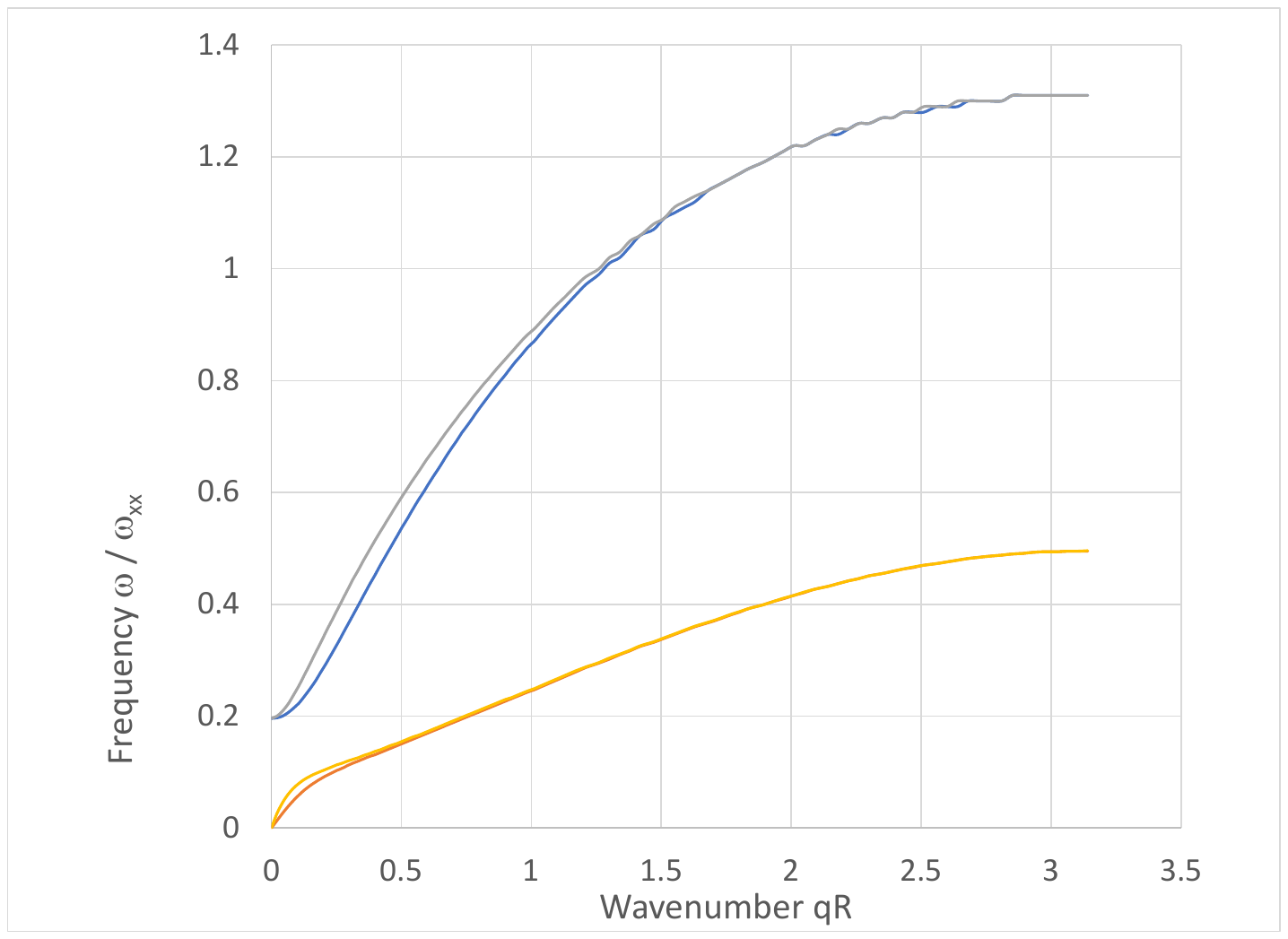}
\caption{
Dimensionless mode frequencies $\omega / \omega_{xx}$ for two strictly one-dimensional chains  where charge movement and atomic polarization can occur only in the x direction
along the chain. The inter-chain Coulomb interaction splits each of the two modes from the one-chain model into even and odd versions, giving 4 modes in total.   Here
 $S_1=0.2,  S_2=0.2, S_3=0, S_4=0, R/R_0=2$.
\label{Modes2Chains_xPolAndCondOnly}
}
\end{figure}

The dispersion energy $\varepsilon _{disp}$ per unit cell is the sum of the modes'
zero point energies, minus the same sum with the inter-chain interactions
turned off:

\begin{equation}
\varepsilon _{disp}\equiv \frac{R}{L}E^{disp}=\frac{1}{2\pi }\int_{-\pi
/R}^{\pi /R}\frac{\hbar }{2}\sum_{i=1}^{4}\left( \omega _{i}\left(
D,q\right) -\omega _{i}\left( \infty ,q\right) \right) dq  \label{Eps_vdW}
\end{equation}

\begin{figure}[tbp]
\includegraphics[width=0.9\textwidth]{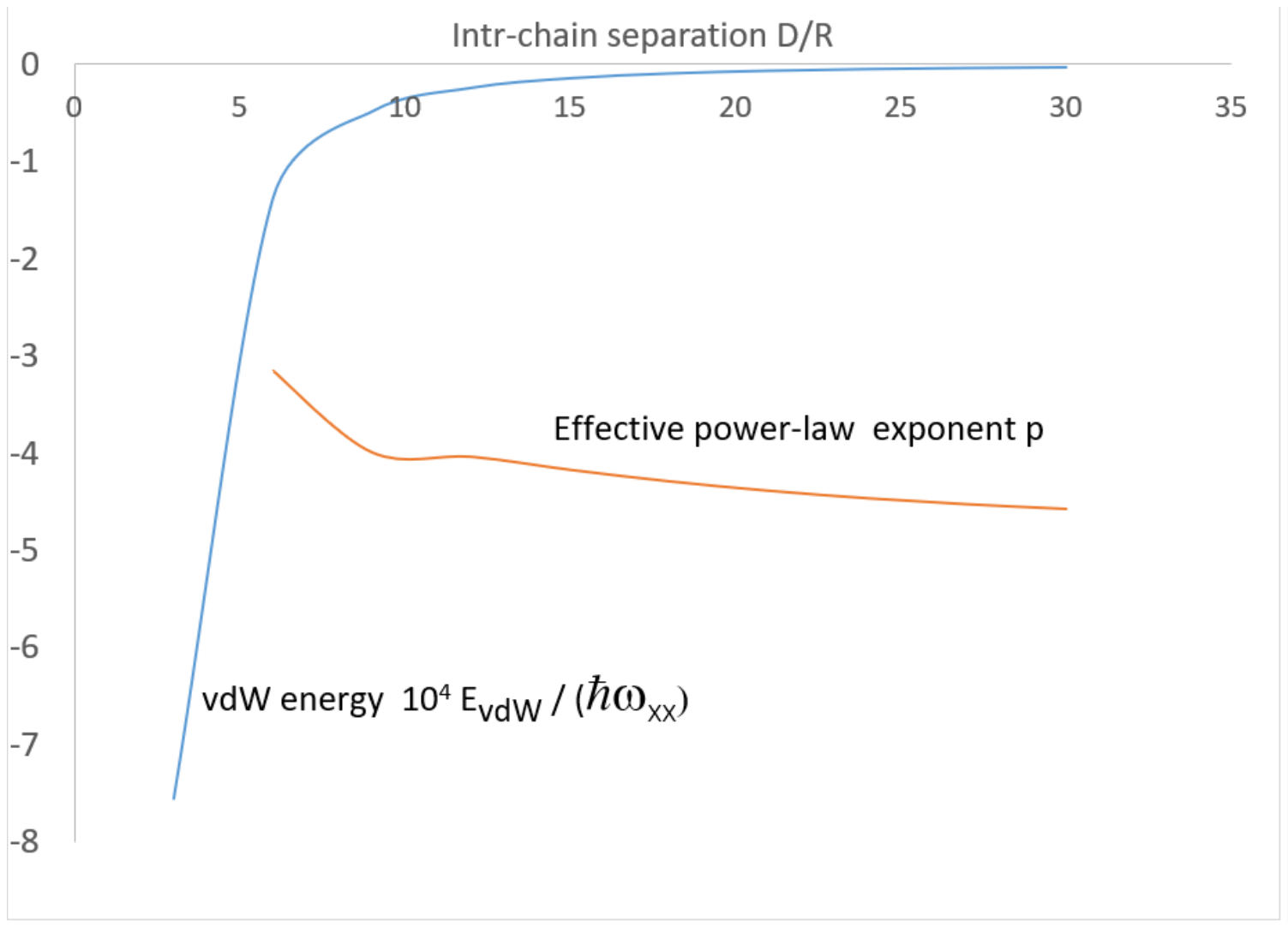}

\caption{
The dispersion energy per atom  from Eq. (\ref{Eps_vdW}) between two strictly 1D  insulating chains with no metallic band, in units of the QHO energy 
$\hbar \omega_{xx}$ of the atoms.    Input parameters were $S_1=0, S_2=0.2, S_3=0, S_4=0$.
 The effective separation-dependent dispersion energy decay  exponent $p=d(ln|E|)/d(lnD)$  is also plotted, showing $p$ approaching $ -5$ at large separations $D$. This corresponds to a separation dependence $\varepsilon=-CD^{-5}$, as predicted for two parallel one-dimensional insulators at virtually any level of dispersion energy theory.
\label{EvdWNoCond}
 }
\end{figure}

\begin{figure}[tbp]
\includegraphics[width=0.9\textwidth]{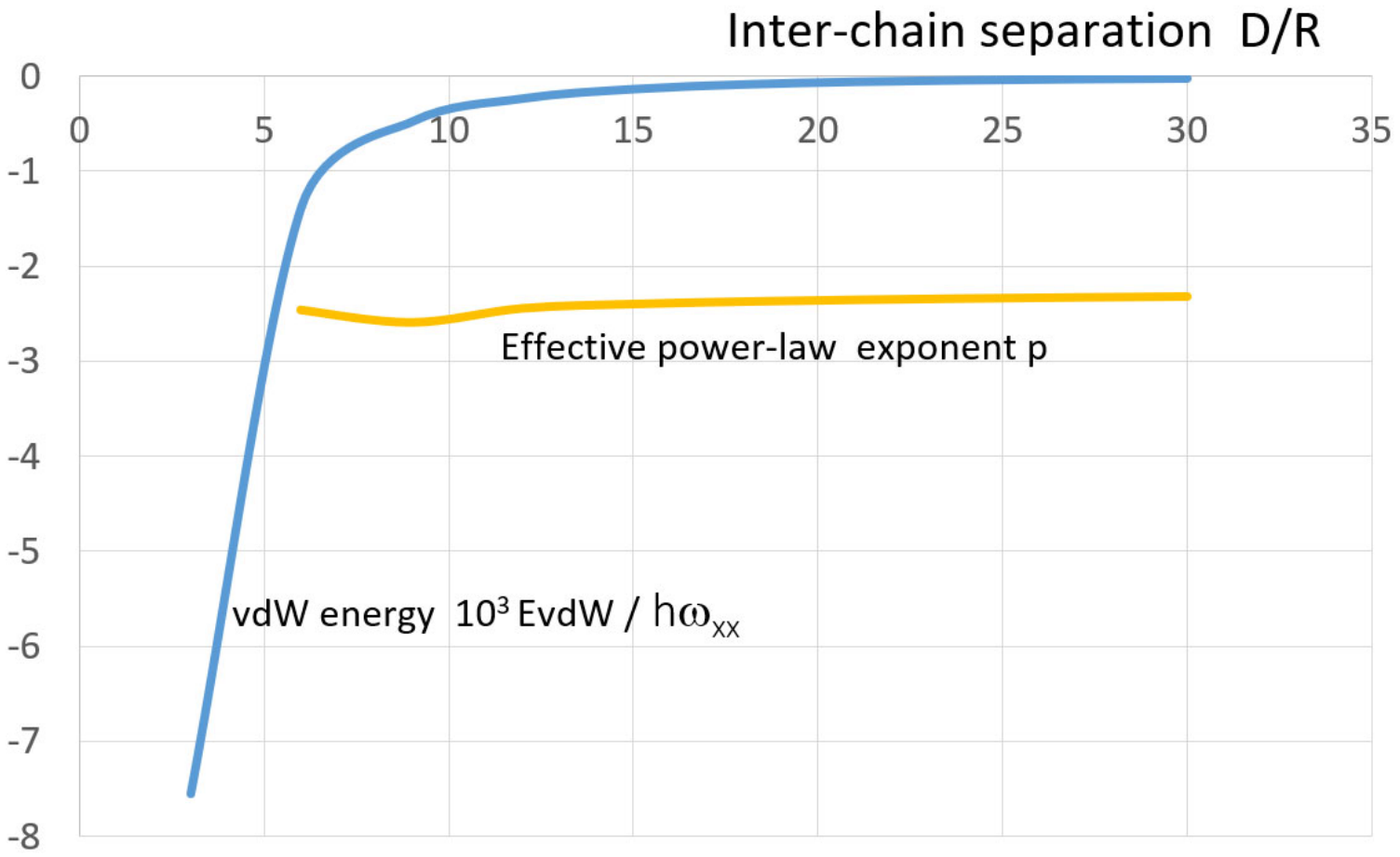}
\caption{
The dispersion energy per atom  from Eq. (\ref{Eps_vdW}) between two  strictly 1D metallic  chains with zero atomic polarizability $\alpha_{xx}=0$, in  units of the QHO energy 
$\hbar \omega_{xx}$ of the atoms.    Input parameters were $S_1=5, S_2=0, S_3=0, S_4=0, R/R_0$=2.
 The effective separation-dependent dispersion energy decay  exponent $p=d(ln|E|)/dlnD$  is also plotted, showing $p \approx -2.3$ at larger separations $D$. This is consistent with the asymptotic  prediction  $\varepsilon\approx L(D)D^{-2}$ from RPA theory, where
 $L(D)$ decays logarithmically with $D$
\label{EvdWNoPol}
 } 
\end{figure}

\begin{figure}[tbp]
\includegraphics[width=0.9\textwidth]{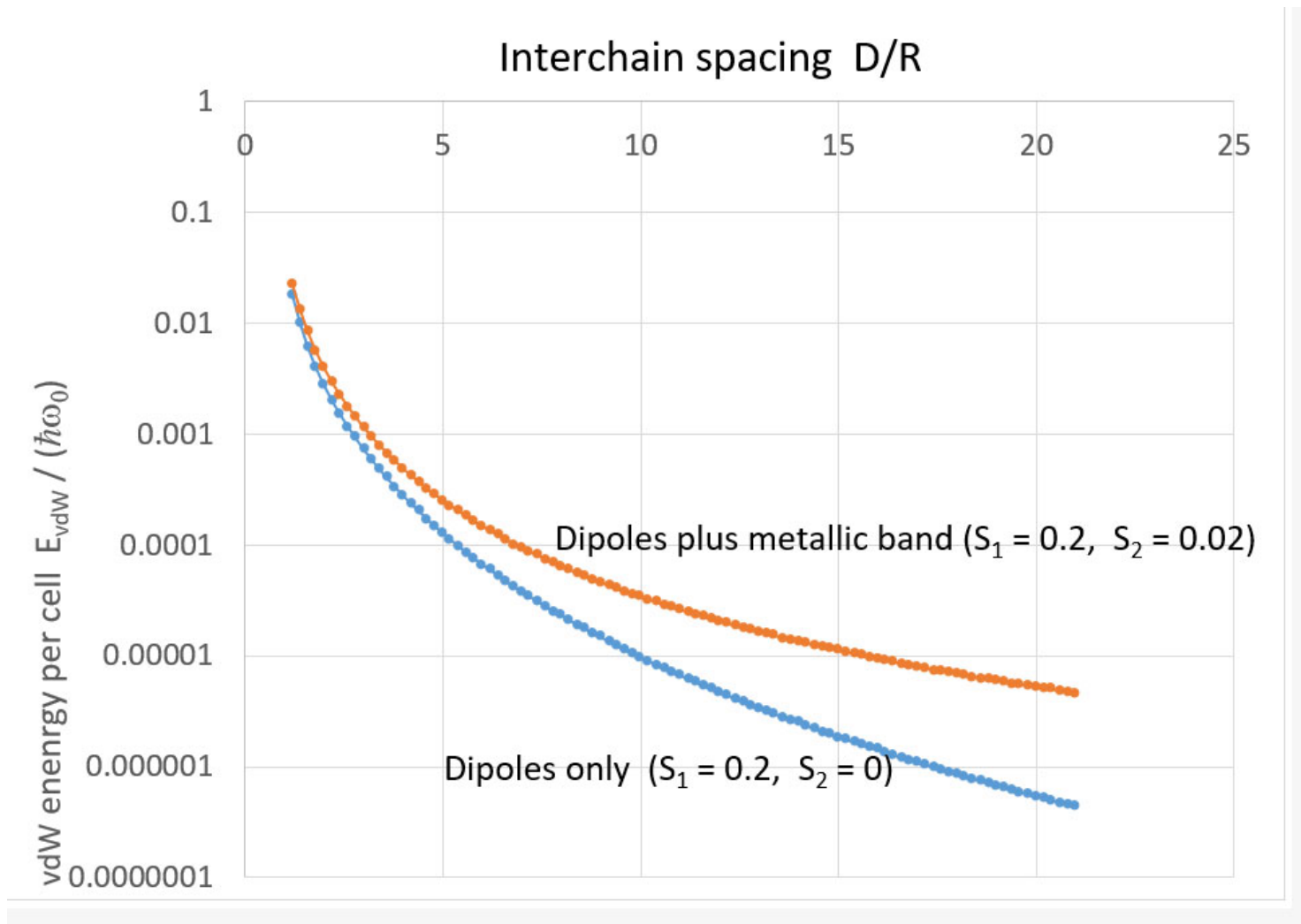}
\caption{
The dispersion energy per atom  from Eq. (\ref{Eps_vdW}) between two  strictly 1D polarizable ($S_2=0.2,S_3=S_4=0$) chains, both with ($S_1=0.02$) and without ($S_1=0$) a weakly metallic band, with $R/R_0 = 2$ .  At larger separations even this weak level of metallicity dominates the dispersion interaction.
\label{EvdWStrictly1D}
 } 
\end{figure}

The sums (\ref{Defn_w00}) etc. and the integral (\ref{Eps_vdW}) were
performed numerically.

 Figs. \ref{EvdWNoCond} - \ref{EvdWStrictly1D} 
 show the dimensionless inter-chain dispersion
energy $\varepsilon _{disp}\;/\;\left( \hbar \omega _{xx}\right) $ as a
function of dimensionless interchain separation $D/R$, for several cases.
Also plotted is the effective exponent $p\left( D\right) =d\left( \ln 
|\varepsilon| \right) /d\left( \ln D\right) $ in an assumed power law 
\begin{equation}
\varepsilon _{disp}=\left( const\right) D^{p}  \label{PowerDecayLaw}
\end{equation}%
\ In the absence of a metallic band, in keeping with Type-A and Type-B
dispersion energy  physics, the value of $p$ is predicted by basic theory to approach the  
pairwise-additive value $p=-5$ at large separations. \ For a metallic
chain the expected value is close to $p=-2$, but this will only occur at
very large separations because theory
\cite{Chang:1971,DaviesNinhamRichmond_1D,dobson-white-rubio} %
shows that for metallic chains the true asymptotic law is (\ref%
{PowerDecayLaw})\ multiplied by a logarithmic factor $\left( \left| \ln
\left( D/const\right) \right| \right) ^{-3/2}$.

Fig \ref{EvdWNoCond} shows numerical results for the dispersion energy from the present theory for the case $%
S_{2}=0.2,\;S_{1}=0,$ which corresponds to a regular MBD-type calculation
with no metallic band, just atoms that are polarizable along the chain
direction. \ As expected, the falloff exponent of $\varepsilon _{disp}$ with $%
D$ does indeed approach $p=-5$ as $D$ increases.

Fig \ref{EvdWNoPol} shows the results from the present discrete MBD-like theory for the
case $S_{2}=0,\;S_{1}=5,\;R/R_{0}=2$ corresponding to strongly metallic
chains with no atomic polarizability. \ The decay exponent $p$ is around $%
-2.3 $ for the largest exhibited values of $D$, and its variation with $D$ is
consistent with a slow decay towards $p=-2$, \ This again is expected from
basic theory, and shows that the present theory, \ although similar to MBD
in that it is based on a discrete atomic description, is capable of
recovering the unusual Type-C decay laws characteristic of low-dimensional
metals predicted by theory
  \cite{Chang:1971,dobson-white-rubio,DrummondNeeds:2007}. The observed slow
 convergence to $p=-2$ is consistent with  the predicted extra logarithmic factor,
\ $E_{disp}\propto D^{-2}\left( \left| \ln
\left( D/D_{0}\right) \right| \right) ^{-3/2}$

\begin{figure}[tbp]
\includegraphics[width=0.9\textwidth]{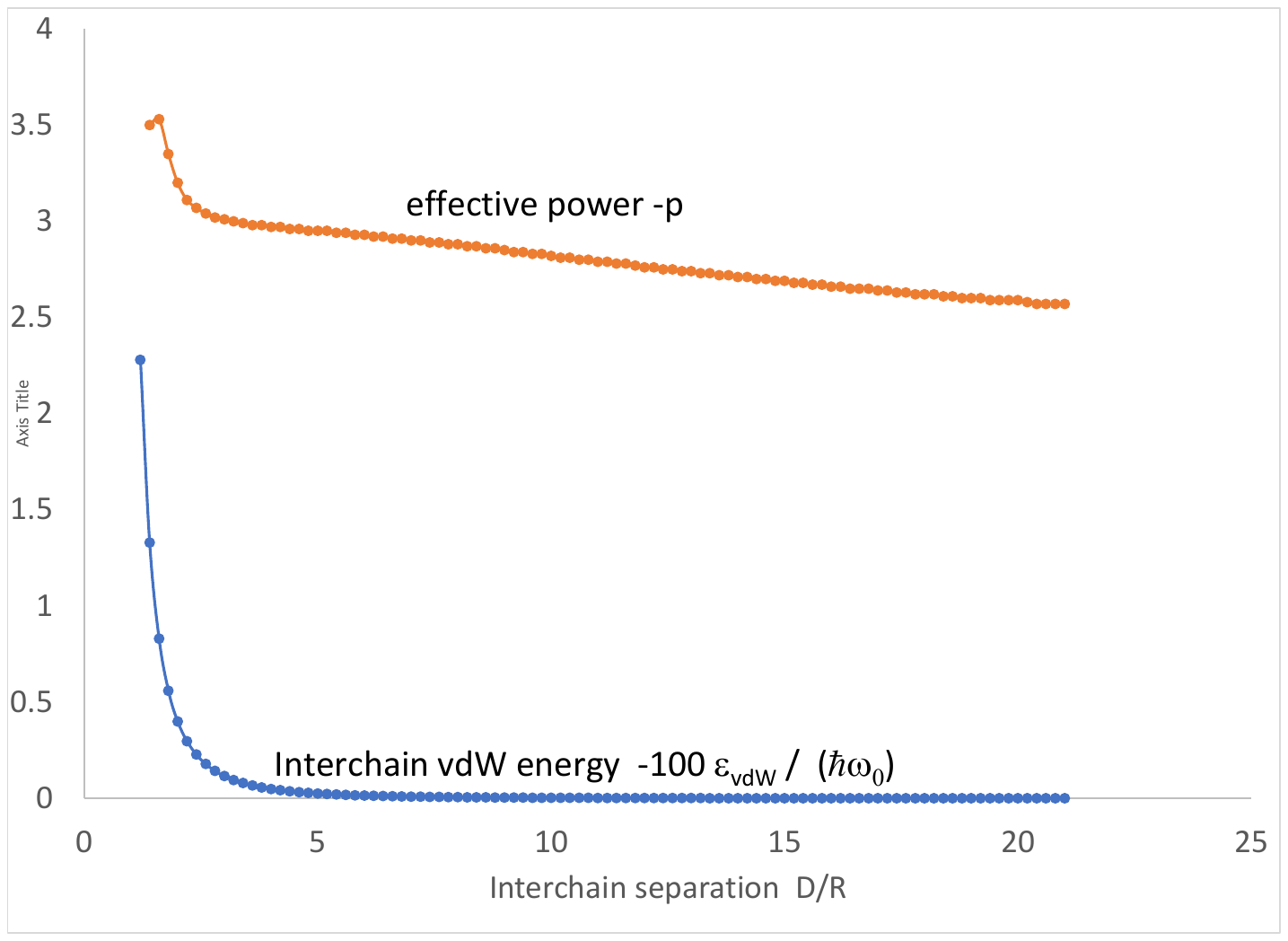}
\caption{
Dispersion energy between  two 1D chains that are strongly polarizable and weakly metallic, with parameters $S_1=0.02, S_2=0.2, S_3=0, S_4=0, R/R_0=2$.  The effective dispersion energy  decay power $p$ eventually approaches the metallic value $p\approx -2$  for very large separations $D\rightarrow \infty$ but takes much more negative values at small separations where the atomic polarizability dominates the physics.
\label{PowerExponentStrbictly1D} 
}
\end{figure}

Fig \ref{EvdWStrictly1D} shows a logarithmic plot of the results $\varepsilon _{disp}\left(
D\right) $ of the present theory for the case $S_{2}=0.2$, \ $%
S_{1}=0.02,\;\;R/R_{0}=2$ corresponding to chains that are each strongly
polarizable and also weakly metallic. \ For comparison the non-metallic case 
$S_{2}=0.2,\;\;S_{1}=0$ is plotted on the same axes. \ At near-contact
geometry ($D/R\approx 2$) the extra energy from adding in the metallic
contribution is only about 25 percent, barely visible in the log plot. \
However at larger separations the metallic \ contribution becomes dominant,
and the effective exponent $p$ of the energy decay moves towards the
asymptotic metallic value $p=-2$. \ (see Fig \ref{PowerExponentStrbictly1D} ).

\section{General case of two coupled monoatomic chains; application to gold 
nanowires}

The semi-analytic solution just presented for inter-chain dispersion interactions
was limited to the case of zero atomic polarizability perpendicular to the
chain. For that case the 2-chain coupled frequencies were obtained from $%
4\times 4$ linear equations that had an analytic solution. \ In the more
general case, where the atoms are polarizable in all directions, the
transverse modes for two chains separated by distance $D$ are no longer
decoupled from the longitudinal modes nor from each other. Thus, in addition
to the couplings from (\ref{Defn_u0D}),(\ref{Defn_T0D}), we have additional
non-zero Coulomb couplings T$_{xy}\left( q,D\right) ,\;T_{xz}\left(
q,D\right) ,\;T_{yy}\left( q,D\right) ,\;T_{yz}\left( q,D\right)
,\;u_{y}\left( q,D\right) ,\;u_{z}\left( q,D\right) .$

These interactions lead to $8\times 8$ equations for the self-sustaining
modes. These require numerical solution for the mode eigenfrequencies $%
\omega _{j}\left( q,D\right) $, followed by an integration $\int \frac{%
\hbar }{2}\omega \left( q,D\right) dq$ \ over the zero-point energies to
obtain the dispersion  energy. We carried this out for the case of two
parallel monoatomic chains of gold atoms.
\subsection{Avoiding double counting for the Au chain}
The model presented above implicitly treated the conduction electrons as a
separate system from the atomic electrons. \ Of course in a gold chain the
outer (mainly s-)\ electrons of each atom are the conduction electrons. The
response of these electrons to ''parallel'' fields\ (i.e. fields directed
along the wire) is treated explicitly above. Thus when considering response
to parallel fields, one should in fact use  the ''atomic'' polarizability
 of the Au$^{+}$  ion,  because the s electrons are treated separately. \
This avoids double counting to a large extent,
 though this issue warrants further investigation in future.
For response to
perpendicular fields, however, our explicit conduction model does not yet
allow for motion of the s electrons perpendicular to the wire, so this must
be included in the ''atomic'' polarizability . \ Thus to avoid severe
double counting we propose ''atomic'' polarizabilities as follows: 
\begin{equation}
\alpha _{xx}=\alpha ^{Au+},\;\;\;\;\alpha _{yy}=\alpha _{zz}=\alpha ^{Au}
\label{AlphasForGodChain}
\end{equation}%
where $\alpha ^{Au+}$ is the (isotropic) polarizability of the Au$^{+}$ion
and $\alpha ^{Au}$ is that of the neutral gold atom. Similarly for the MBD
harmonic oscillator frequencies \ we take%
\begin{equation}
\omega_{xx}=\omega ^{Au+},\;\;\;\;\omega _{yy}=\omega _{zz}=\omega ^{Au}
\label{OmegasForGoldChain}
\end{equation}

Similar choices will be appropriate for metals in general, with specific
account of the number and direction of the metallic bonds to inform the
choice of atomic vs. ionic input data.

\subsection{
Parameters for a gold monowire:
  geometry, atomic polarizabilities and frequencies}
Jariwala et al. \cite{jariwala} have \ performed DFT\
calculations of equilibrium geometries and electronic Bloch bandstructures $%
\varepsilon \left( k\right) $ for several types of gold nanowire. For the
monoatomic chain they obtained an equilibrium inter-atom spacing$\;$(their
Table\ 1)%
\[
R=2.52\; Angstrom=\;\frac{2.52}{0.5292}a_{B}=4.76\;\;a_{B}\;. 
\]
By measuring the gradient of their Bloch bandstructure ( $\varepsilon _{k}$
vs. $k$) plot (their Fig. 4) we obtained\ for the conduction electrons\
\thinspace 
\[
v_{F}=\frac{\partial \varepsilon _{k}}{\hbar \partial k}|_{k_{F}}=7.\,72%
\times 10^{5}\;m/s 
\]%
For comparison, the free-electron 1D Fermi velocity at a density of 1
electron per lattice cell ($R=2.52\;$\AA) is 
\[
v_{F}^{free}=\frac{\hbar \pi }{2Rm}=\frac{\left( 1.05\times 10^{-34}\right)
\pi }{2\left( 2.52\times 10^{-10}\right) \left( 9.11\times 10^{-31}\right) }%
=7.18 \times 10^{5}\;\;m/s
\] 
This is quite similar to the band-theory value obtained just above,
confirming that 1D gold is a nearly-free-electron metal, at least  within
independent-electron theory. 

We used literature sources for the static polarizability of the Au
atom and of the Au$^{+}$ cation. Table 1 shows these values in atomic units \ The derivation of the Au$^+$ oscillator frequency $\omega _{0}$ is discussed below

\bigskip 
$
\begin{array}{|c||c|c|c|c|}
\hline
\text{Atomic units}  \;(a_{Bohr}^{3}\text{, Hartree}/\hbar ) & \alpha _{0}\left(
0\right) \;Au & \omega _{0}\;Au & \alpha _{0}\left( 0\right) \;Au^{+} & 
\omega _{0}\;Au^{+} \\ 
\hline \hline
\text{Artiukhin \& Buchachenko \cite{artiukhin}} & - & - & 13.36 & 0.409\;\;(J:\approx
0.7) \\  
\hline
\text{Original MBD \cite{mbd,mbdrs}} & 36.5 & 0.298 & - & -
 \\ 
\hline
\end{array}%
$
\bigskip

For the Au$^+$ ion, Artiukin and Buchachenko \cite{artiukhin} only report polarizability 
$\alpha_0^{\rm Au+}$. Here, the corresponding oscillator frequency $\omega_0^{\rm Au+}$ 
is estimated exploiting the f-sum rule. In fact, conservation of the number of electrons implies that  
$\alpha_0 \omega_0^2 = \alpha_0^{\rm Au+} (\omega_0^{\rm Au+})^2 + 1$, 
so that, exploiting the MBD parameters for neutral Au, one obtains $\omega_0^{\rm Au+}=0.409\, a.u.$.

Within our scheme defined in Eq. (\ref{AlphasForGodChain}), with the $\alpha
_{0}^{Au+}$ value from Artiukhin \& Buchachenko (\cite{artiukhin}, A\&B) used for 
$\alpha _{xx}$, we have $\alpha _{xx}R^{-3}=$ $13.6$/$\left( 4.76\right)
^{3}=0.126\,16<0.207\,94$. Thus according to (\ref{StabilityCond_x}), the
single-chain parallel-polarization mode is stable even without any damping/softening  
of the point-dipole Coulomb interaction at short range (See the following Paragraph). 
The single-chain perpendicular modes are also stable without any Coulomb damping, 
since  with original MBD data\cite{mbd,mbdrs} for 
$\alpha_0^{Au}$ we have $\alpha _{yy}R^{-3}=\alpha _{zz}R^{-3}=$ 
$36.5/\left( 4.76\right) ^{3}=\allowbreak 0.338\,4\leq 0.554\,6$.
Thus  the stability criterion (\ref{StabilityCond_y_z}) \ is satisfied. \ 

\subsection{Coulomb damping to allow for finite radii of the atomic orbitals }
In the analytic working above we treated  dynamic dipoles and charges as point objects, for the purpose of  calculating the Coulomb interaction between them.  In fact it is more realistic to smear out these charge distributions to reflect the finite radius of the atomic orbitals that contribute to them.  The Coulomb interaction between  these smeared  charge distributions will here be termed the "damped Coulomb interaction"

Here we chose Gaussian Coulomb damping parameters as follows: the Coulomb interaction between two unit charges (e) situated at $\vec{R}_I$, $\vec{R}_J$ was renormalized at short range according to the following expression
\begin{equation}
w_{damp}(|\vec{R}_I - \vec{R}_J|)=\frac{erf(|\vec{R}_I - \vec{R}_J|/\sigma_{IJ})}{|\vec{R}_I - \vec{R}_J|}\,.
\end{equation}
Here $\sigma_{IJ}=\sqrt{\sigma_I^2 + \sigma_J^2}$, where $\sigma_I=(\sqrt{2/\pi}\alpha_{0,I}/3)^{1/3}$ for QHO's ($I$ is the atomic index), according to the MBD approach\cite{proof,science}.
This damping effectively accounts for the finite width of the Gaussian charge distribution which naturally emerges from QHO's,
thereby reducing the Coulomb interaction when two QHO's overlap.
To extend the above expression to metallic charges, we selected 
$\sigma_J = R/2 =1.26$ \AA, 
whenever one of the two interacting objects
is a metallic electron. This parameter was selected  
because the s-electron clouds are presumably in contact at the  equilibrium inter-atomic spacing $R=2.52$ \AA, so that $R/2$ is a rough estimate of the effective s orbital radius {\it in situ}.   The on-site charge self-interaction radius was also chosen to be
  $R_0 = R/2$ for the same reason.  This value ensures stability of the single-chain 1D
plasmon mode as per Eq.  (\ref{StabiityCond_plasmon})
 The charge-dipole and dipole-dipole interactions are then obtained by
 differentiating  $w_{damp}$ with respect to $\vec{R}_{I,J}$
and selecting the appropriate $\sigma_{I,J}$ parameters.

\subsection{Type-A polarizability scaling based on atomic volume }
In standard MBD  calculations the input atomic polarizabilites are often reduced from the free-atom values by considering Hirshfeld volumes of the atoms \emph{ in situ} \cite {ts}.  This is sometimes termed "Type A non-additivity" \cite{dobson-abc}. We applied this procedure to all the calculations below.   For the gold chain we found only a small a reduction from this procedure.

\subsection{The contact regime of  inter-metallic bonding in general }
 The semi-local DFT  calculation by Jariwala et al.\cite{jariwala} showed that two closely spaced
monoatomic Au chains undergo a metallic bonding process, with one stable
structure being a twinned ''ladder''. \ The equilibrium inter-chain binding
separation for a gold ladder was found to be $D=D_{0}=2.52$ \AA. 

This is an example of the general phenomenon of inter-metallic bonding at contact.  The energy and geometry at inter-metallic contact are already well described by semi-local density functional theory (DFT), because the Kohn-Sham kinetic energy dominates the binding via orbital overlap.    Hence theories such as MBD, which calculate a highly nonlocal inter-atomic electron correlation energy component, are not needed here, giving only a minor energy contribution if correctly applied. 

The nn-MBD+C approach can indeed  describe the dispersion energy in this overlapped regime, but in the present  case of gold chains this requires a 
version of nn-MBD+C in which the electronic  KS response $ \chi_0$ allows electrons
 to tunnel both between the chains and along them  (see Eq. (\ref{GeneralNN_Chi0Ansatz})), whereas the version developed  above only permits inter-atomic charge movement  along each chain.   The more complete model of $\chi_0$  
can be constructed from a tight-binding approach, but we leave this to a possible  future investigation because, as just explained, it is unlikely to make a significant difference to the total energy in the metallic binding regime.   

Conversely, success of an MBD-type theory (when coupled appropriately with DFT)  in predicting the metallic contact energy and geometry  is not a meaningful test of the nonlocal MBD contribution  - the success is due to the DFT component.

Because of the above considerations, and based on the magnitude of the  Type-C corrections found below for geometries just outside contact, we expect the Type-C metallic corrections at contact will be less than 1\% of the large binding energy.

Throughout this paper we will therefore restrict calculations to the non-overlapped regime where the nonlocal correlation energy provided by MBD-type schemes is the dominant attractive energy.  For the present case of monoatomic gold chains we calculate the energy for separations down to $D = 3 \AA $, just short of the equilibrium spacing of $2.52 \AA$.

The non-contact regime is important in general to model self-assembly, docking, catalysis and reaction energy barriers, and in molecular dynamics calculations.

\subsection{Charge and polarization modes of  two parallel gold chains}
\begin{figure}[tbp]

\includegraphics[width=0.9\textwidth]{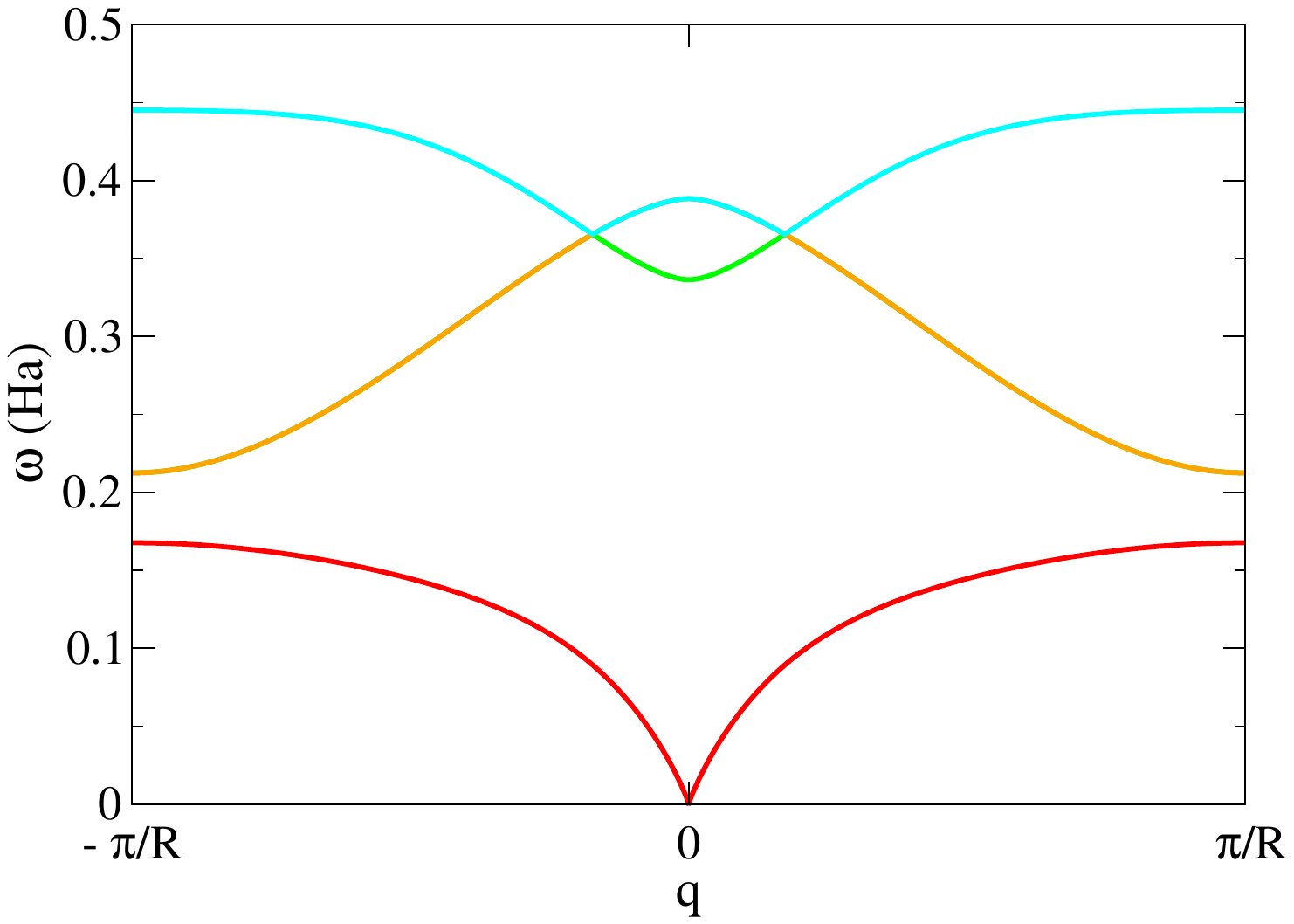}
\caption{
nn-MBD+C mode frequencies vs. wavenumber for a single chain of Au atoms with  Gaussian Coulomb damping.  Only three curves are present because the y-directed and z-directed polarization waves are degenerate. Atomic units are used.   The gapless 1D plasmon mode seen here is not present in regular MBD theory.
\label{ModeFreqs_1Au_chain}
 } 
\end{figure}

Figure \ref{ModeFreqs_1Au_chain}  shows the nn-MBD+C modes of a single Au chain with Gaussian Coulomb damping.  The lowest-frequency mode (gapless 1D plasmon) appearing here is not present in regular MBD theory.

\begin{figure}[tbp]

\includegraphics[width=0.9\textwidth]{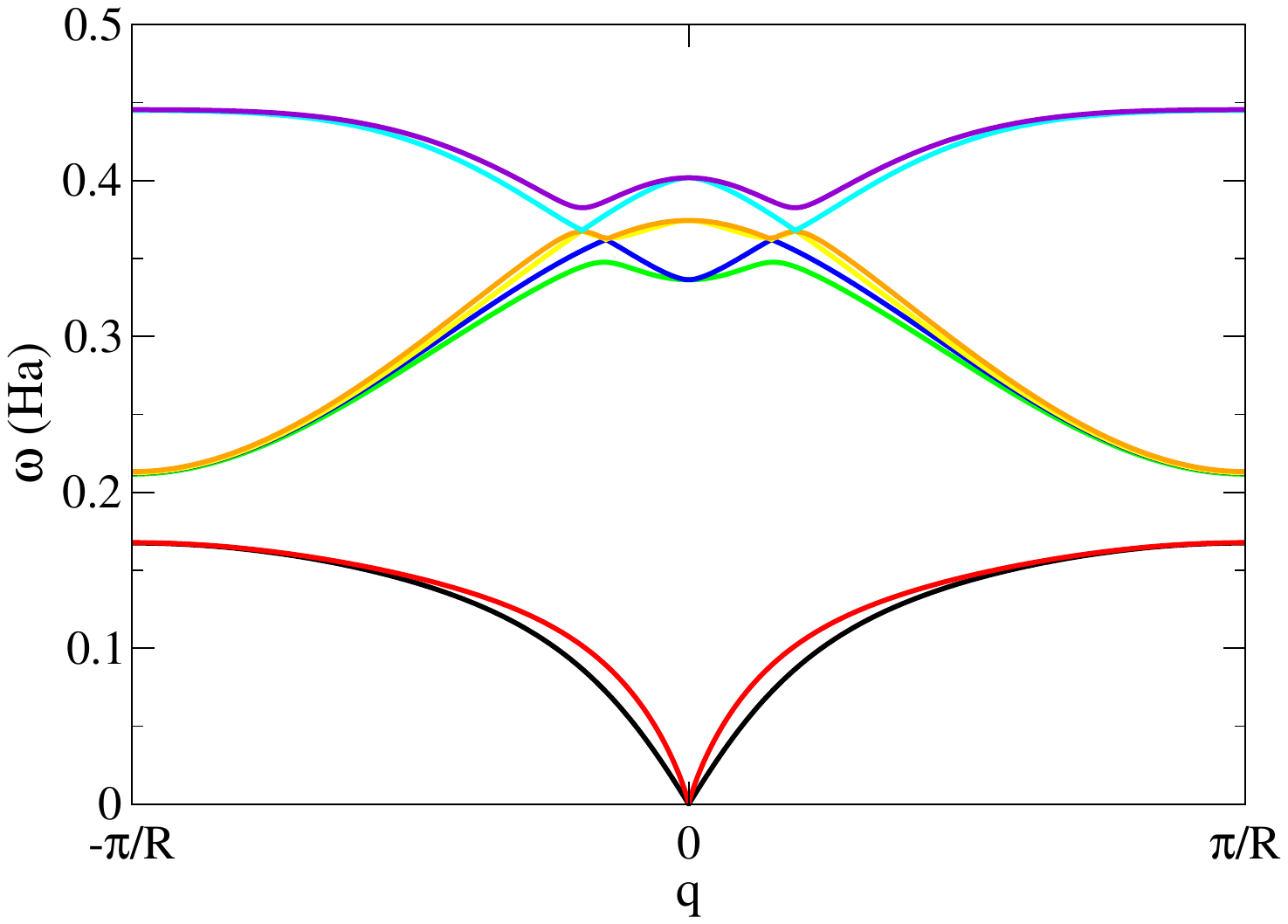}
\caption{
nn-MBD+C mode frequencies versus wavenumber for two parallel gold chains with inter-chain separation $D=6$ \AA,  with Gaussian Coulomb damping.
\label{ModeFreqs_2Au_chains_6A}
}
\end{figure}

Figure \ref{ModeFreqs_2Au_chains_6A}  shows the nn-MBD+C modes of two parallel Au chains separated by a distance $D=6$ \AA,  with Gaussian Coulomb damping.
  The mode frequencies were obtained by numerical evaluation  of the eigenvalues of an $8\times 8$ matrix for each wavenumber $q$, as explained above.  Each of the single-chain modes from figure 
\ref{ModeFreqs_1Au_chain} is now split into even and odd versions as expected from the mirror-symmetric geometry.

\begin{figure}[tbp]

\includegraphics[width=0.9\textwidth]{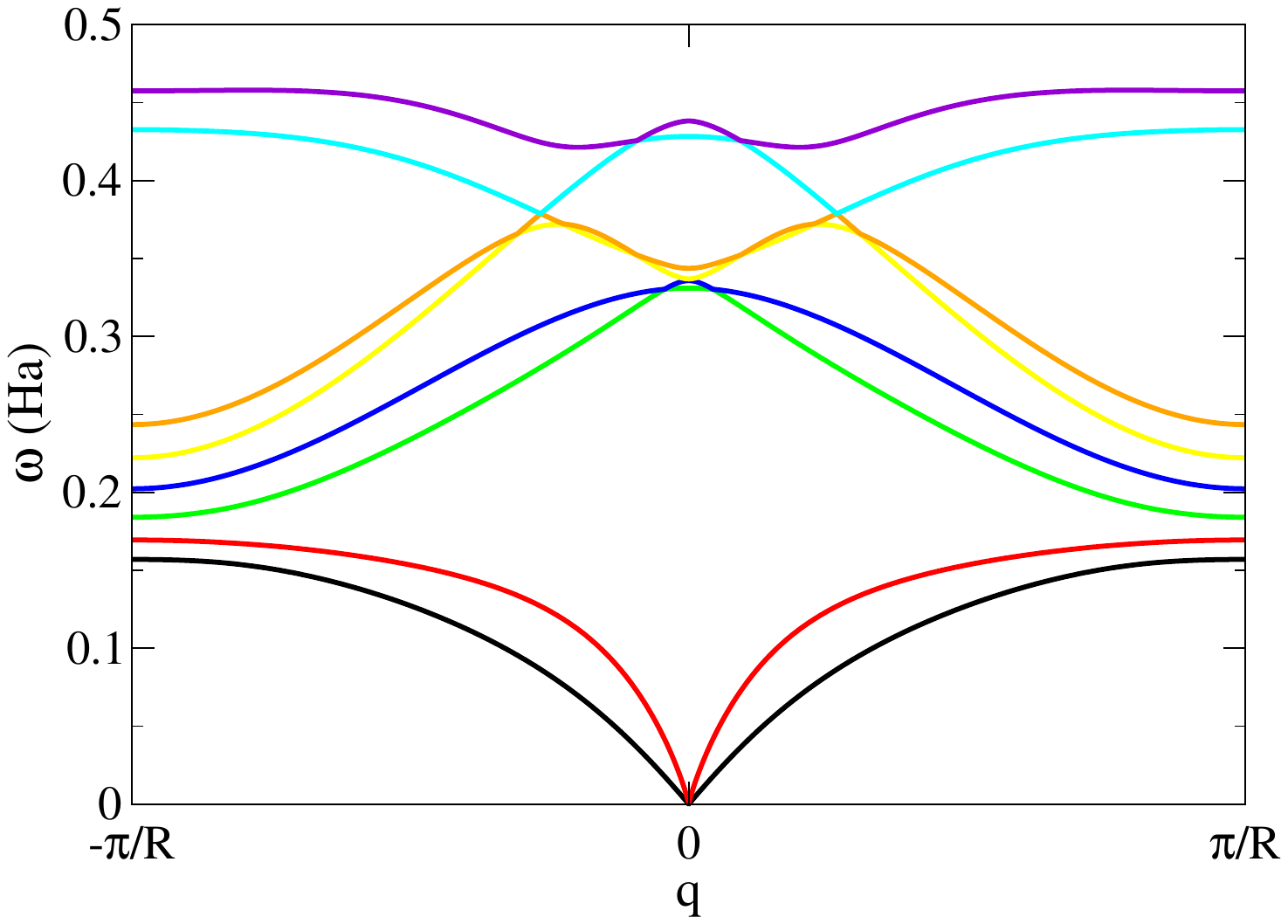}
\caption{
nn-MBD+C mode frequencies versus wavenumber for two parallel gold chains with inter-chain separation $D=3$ \AA,   with Gaussian Coulomb damping.
  This geometry is close to the  metallic bonding regime.
\label{ModeFreqs_2Au_chains_3A}
}
\end{figure}

Figure \ref{ModeFreqs_2Au_chains_3A}  shows the MBD+C modes of a two parallel Au chains separated by a distance $D=3$ \AA,  with Gaussian Coulomb damping.
 All modes are still stable, even though this geometry has the chains close to metallic contact.

\subsection{\ Parallel gold chains: dispersion energies}

The dispersion  energy was obtained by numerical integration of the
\ mode zero-point energies over wavenumer $q,$ as per Eq. (\ref{Eps_vdW}). 
We sampled the Brillouin zone with 2000 q-points, which provided sufficient
convergence up to the largest interchain distances studied here.

\begin{figure}[tbp]
\includegraphics[width=0.9\textwidth]{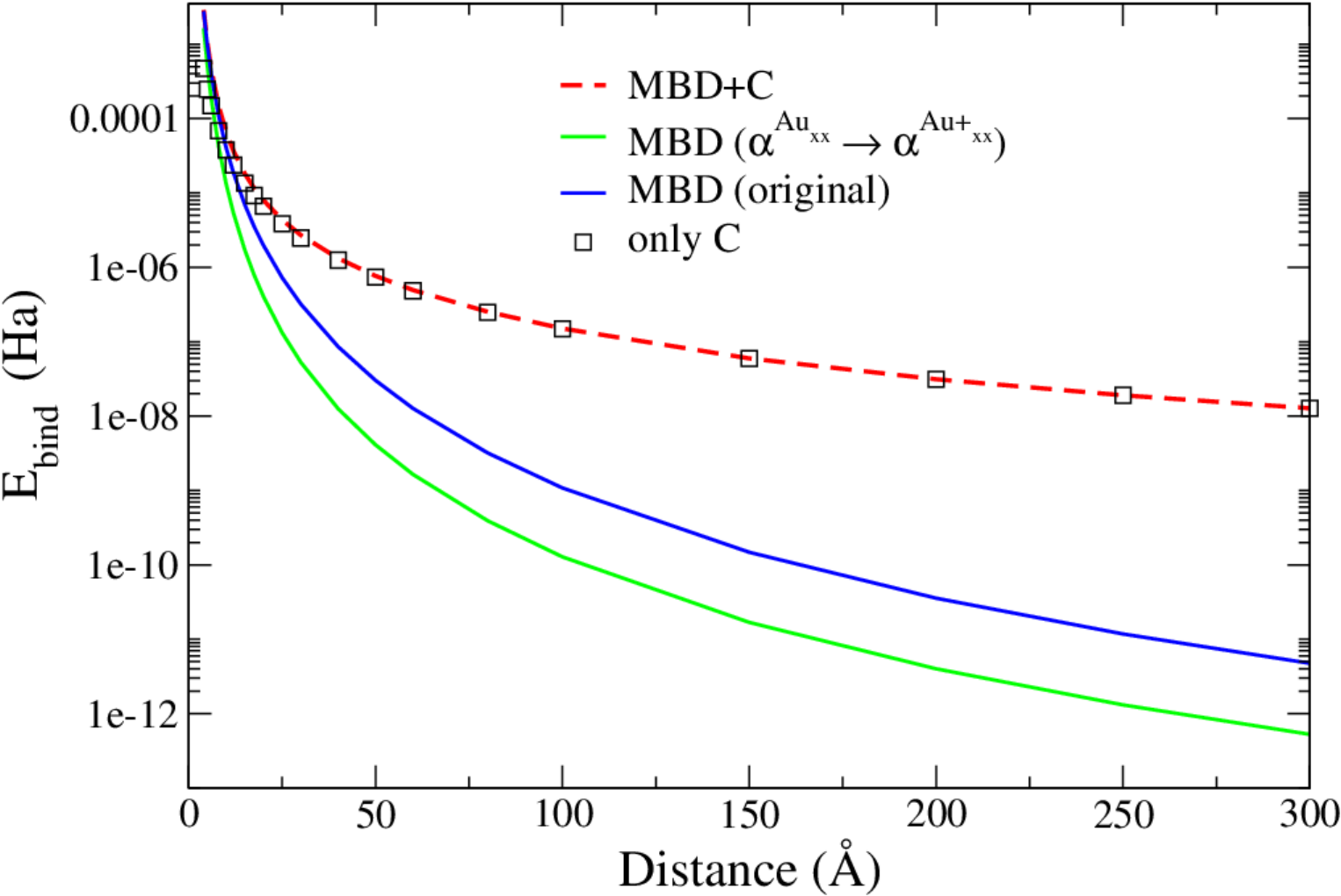}

\caption{
Dispersion energy per atom  between parallel gold chains as a function of inter-chain separation  from MBD and nn-MBD+C, with Coulomb damping.
The contribution from conduction electrons only is also given (only C)
 \label{LogEnergy_2Au_chains}
}
\end{figure}

\begin{figure}[tbp]
\includegraphics[width=0.9\textwidth]{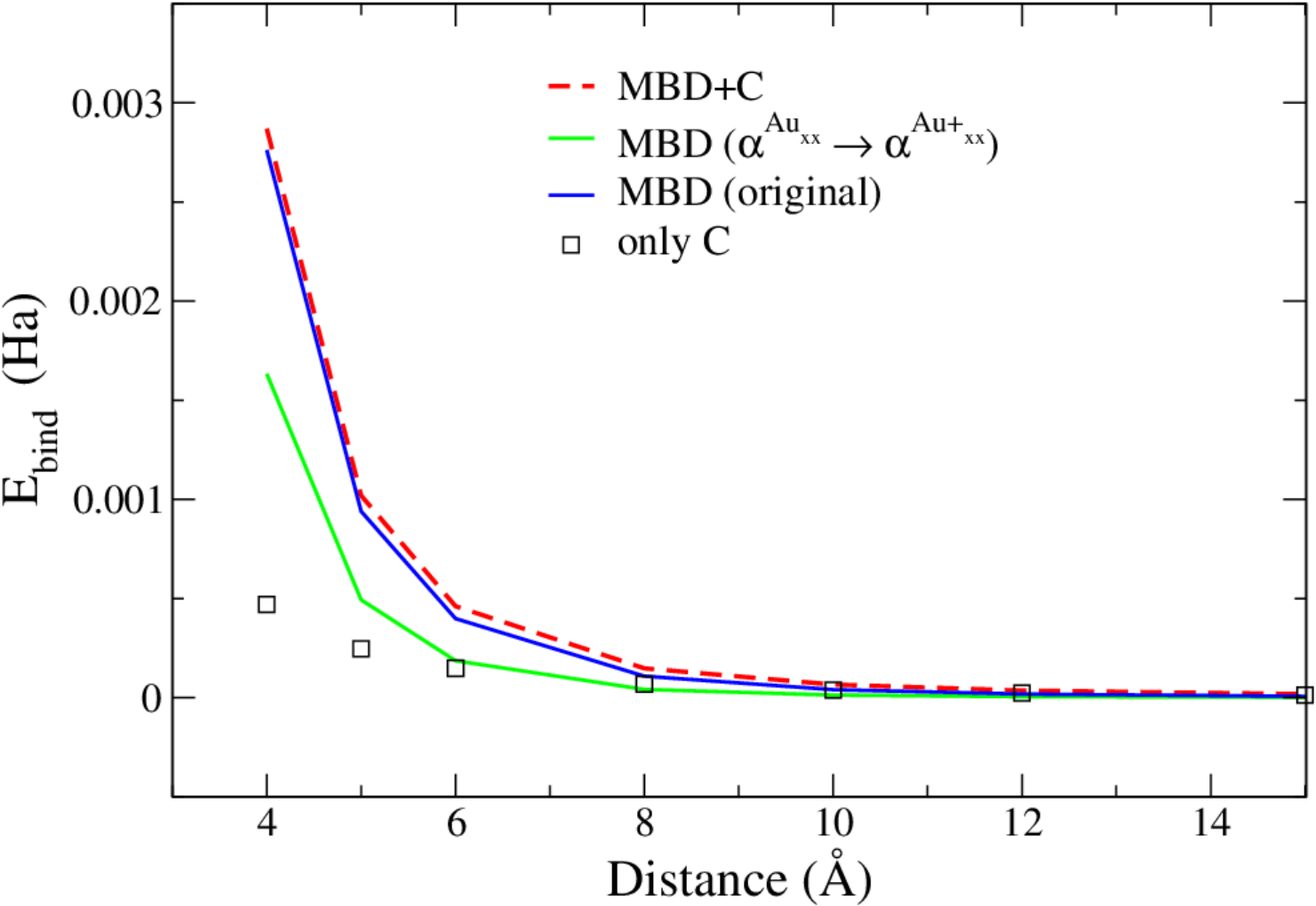}
\caption{
Dispersion energy per atom  between parallel gold chains from MBD and nn-MBD+C, with Coulomb damping, 
showing detail for smaller interchain distances  $D$. 
At near-contact separations nn-MBD+C gives slightly more attractive dispersion energy than regular MBD (by  4\% at $D=4$ Angstrom). This contrasts with Fig 
\ref{LogEnergy_2Au_chains}, which shows that, at larger separations, the MBD+C interaction exceeds that from MBD by a large  factor, in agreement with high-level asymptotic theory.
\label{Energy_2Au_chains}
}
\end{figure}

Figure \ref{LogEnergy_2Au_chains}  shows the dispersion energy $-E_{disp}\left( D\right) $ as a function of separation $D$ for larger $D$ values, from MBD and nn-MBD+C. 
 The contribution from conduction electrons only is also computed for reference. As expected from the analytic solutions to  the 1D model given  above, the metallic Type-C contribution
dominates at large separations, with atomic polarization being more important
at smaller $D$ values - see Fig. \ref{Energy_2Au_chains} for more detail there.

 Fig. \ref{Energy_2Au_chains} shows that,  at narrow separations $D$, nn-MBD+C gives results quite close to the original MBD (within $4\%$ at $D=4$ Angstrom), despite the failure of the latter to obtain the correct asymptotic power law.   This relative smallness of the near-contact   Type-C contribution is possibly due to the large polarizability of the Au core.  Indeed, the conduction electrons here are a small fraction of the polarizable electrons present.  Larger Type-C effects are seen  for the carbon-based systems studied below.

\begin{figure}[tbp]
\includegraphics[width=0.9\textwidth]{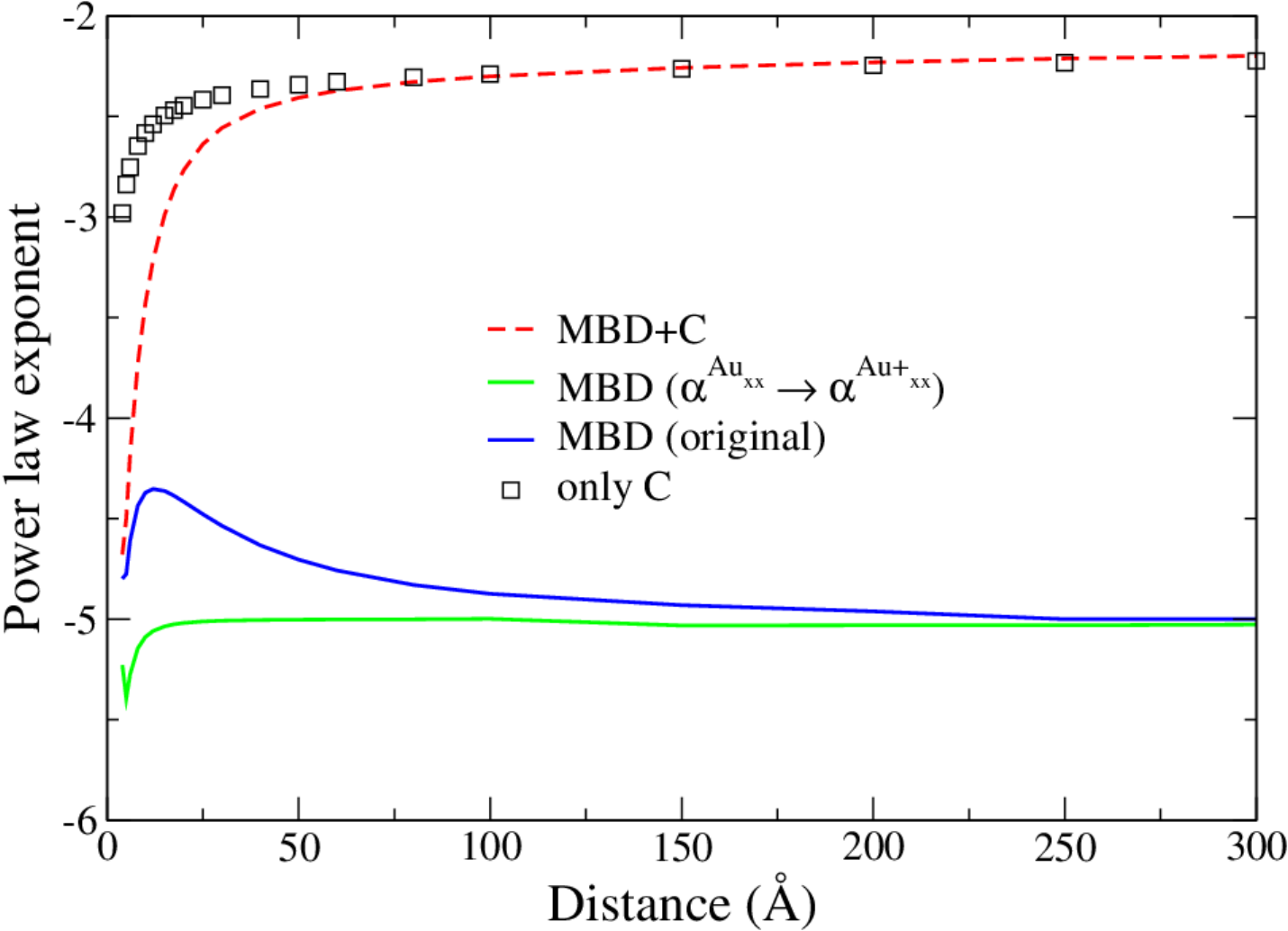}
\caption{
Effective power law decay exponent 
 $p=d \ln  |E\left( D\right)| /d (\ln D)$
of the dispersion interaction between two parallel gold chains. 
in various theories.
\label{Eff_power_2Au_chains} 
}
\end{figure}

 Figure \ref{Eff_power_2Au_chains}  shows the effective dispersion energy  power law exponent
 $p=d \ln \left|E\left( D\right) \right| /d (\ln D)$ for the decay of the dispersion energy
 $E\left( D\right) $ between two gold chains as a function of chain separation $D$. 
\ At the larger separations it shows $p\approx -2.3$, consistent with
Type-C dispersion interactions as discussed above, while for smaller distances more
negative values are obtained, similarly to regular MBD.

These results show that our nn-MBD+C scheme smoothly connects conventional and
metallic dispersion energy regimes, as the inter-chain separation $D$ is increased. \
nn-MBD+C is only slightly more intensive numerically than regular MBD, here
involving $8\times 8$ matrices rather than $6\times 6$.

\section{Lithium-doped graphene}
We consider two parallel sheets of graphene with Li atoms placed  outside this graphene bilayer on a 2$\times$2 superlattice, so that there is one Li atom for every 8 C atoms.  The Lii nuclei are placed   $1.79$ Angstrom  outside the planes containing the C nuclei.  
In all of the calculations reported below for this system
we implemented nn-MBD+C using Gaussian Coulomb damping and Hirshfeld-volume-based polarizability renormalization.

\subsection{Li@graphene in original MBD}
 The original MBD method takes all entities as neutral atoms.  If the Li atoms are taken not to be ionized then we have two non-metallic sheets, whose dispersion interactions we evaluated using the original MBD approach.    Since the bilayer is periodic in the  $xy$ plane with  16 atoms in a  cell, we used a Fourier transform technique to implement MBD, with the dispersion energy emerging as a  numerical integral of eigenfrequencies over a two-dimensional Brillouin zone.   Some details about this type of Fourier transform are also needed to implement nn-MBD+C, and are given the next Subsection.
 In Figs \ref{Log_plot_energies_Li_gaphene} and \ref{Linear_plot_energies_Li_graphene} the MBD  dispersion energy between the layers (assuming neutral atoms)  is plotted vs. distance $D$ between the planes of C atoms (solid blue line).   We also calculated the interaction between plain graphene sheets without lithium (open squares).  The MBD energy without Li is very close to the result with Li f or near-contact separations, but  falls away more sharply with distance $D$.  

\subsection{nn-MBD+C for fully ionized Li@graphene}
In reality Li will ionize, causing electron doping and creating two metallic graphene sheets. 
Assuming fully ionized Li, we used the new nn-MBD+C theory to evaluate the inter-sheet dispersion interaction. The response of the dopant electrons to fields parallel to the plane was treated via the 2D Bloch electron model inherent in nearest-neighbor  MBD+C theory as expounded above (Eq (\ref{Chi0Ansatz_graphene})).    To establish the value of the constant $B$ in (\ref{Chi0Ansatz_graphene})  we Fourier-transform this equation.   The graphene lattice has two basis atoms per direct unit cell, here labelled $p=1,2$, and  in general an atom in a lattice with a basis can be labelled $(p,\vec{R})$ where $\vec{R}$ is the origin of its unit cell.   The atom's spatial position is then $\vec{r}_{p,\vec{R}} = \vec{R}+\vec{x}_p$ where $\vec{x}_p$ is the atom's position relative to the origin of its cell. 
For lattice function $g_{p\vec{R}}$ we define the discrete Fourier
transform as  %
$
g_{p}\left[ \vec{q}\right]  =\sum_{\vec{R}}g_{p\vec{R}}\exp \left( i\vec{q}%
\cdot \vec{r}_{p\vec{R}}\right)  
$.
For a 2-point translationally invariant function
 $f_{pp^{\prime }\vec{R}-\vec{R}^{\prime }}$ 
\ we define the discrete Fourier transform as 
\begin{equation}
f_{pp[}\left[ q\right] =\sum_{\vec{X}}f_{pp^{\prime }\vec{X}}\exp \left( i%
\vec{q}\cdot\left( \vec{X}+\vec{x}_{p}-\vec{x}_{p^{\prime }}\right) \right)
=\sum_{\vec{R}-\vec{R}^{\prime }}f_{pp^{\prime }\vec{X}}\exp \left( i\vec{q}%
\cdot\left( \vec{r}_{p\vec{R}}-\vec{r}_{p^{\prime \vec{R}^{\prime }}}\right)
\right)  
 \label{Defn_discrete_FT_2point_fn}
\end{equation}%
where $\vec{X}\equiv \vec{R}-\vec{R}^{\prime}.$

Then the components of the discrete Fourier transform $\chi_{0pp'}[\vec{q}]$   of\ (\ref{Chi0Ansatz)_1Dchain}) 
using the definition (\ref{Defn_discrete_FT_2point_fn}) are%
\begin{equation}
\frac{\omega ^{2}}{B}\chi _{011 }\left[ \vec{q},\omega \right]  =%
\frac{\omega ^{2}}{B}\chi _{022 }\left[ \vec{q},\omega \right]  = 1
\label{Realspace_chi0_pp}
\end{equation}
\begin{equation}
\frac{\omega ^{2}}{B}\chi _{012 }\left[ \vec{q},\omega \right] =-%
\frac{\omega ^{2}}{B}\chi _{021}\left[ \vec{q},\omega \right]^* =
\frac{1}{3}\exp (i\vec{q}\cdot\vec{X}_{K})-\frac{1}{3}\exp (i\vec{q}\cdot\vec{X}%
_{L})-\frac{1}{3}\exp (i\vec{q}\cdot\vec{X}_{M})
\label{Realspace_chi0_ppprime}
\end{equation}
Here $\vec{X}_1, \vec{X}_2, \vec{X}_3$ are the vectors pointing from a $p=1$ atom to its nearest neighbors on the graphene lattice. These expressions for $\chi_0[\vec{q},\omega]$ are complex. \ However for comparison with the long-wavelength ($\vec{q} \rightarrow \vec{0} $) metallic response from Bloch electron theory  we only need the total charge per cell induced by an applied potential that is essentially constant across another cell.   This is described by the following  contraction:
\begin{eqnarray*}
\frac{\omega ^{2}}{B}\sum_{pp^{\prime }}\chi _{0pp^{\prime }\text{ }}\left[ 
\vec{q},\omega \right] &=&2\left[ 1-\frac{1}{3}\cos \left( \vec{q}\cdot\vec{X}%
_{K}\right) -\frac{1}{3}\cos \left(\vec{q}\cdot\vec{X}_{L}\right)-\frac{1}{3}\cos \left(\vec{q}\cdot%
\vec{X}_{M}\right)\right]  \\
&=&\frac{1}{3}\left[ \left( \vec{q}\cdot\vec{X}_{K}\right) ^{2}+\left(\vec{q}\cdot\vec{X}%
{L}\right)^{2}+\left(\vec{q}\cdot\vec{X}_{N}\right)^{2}\right] +O\left( q^{4}\right)
\;\;as\;q\rightarrow  \vec{0} \\
&=& \frac{1}{6}a^2q^2 +O(q^4)
\end{eqnarray*}
where the last two lines used the specific geomtry of graphene, and  $a = 2.46$ Angstrom  is the length of the direct basis vectors in graphene. The constant $B$  must be chosen to make  this discrete transform agree  with the continuous transform  $\chi_0(q,\omega)$ of the known 2D Bloch electron response from Eq (\ref{Chi0_2Dmetal}):
\begin{eqnarray}
\frac{B}{6\omega ^{2}}a^{2}q^{2} &=&\Omega _{cell}\frac{N_{s}N_{v}v_{F}k_{F}%
}{4\pi \hbar }\frac{q^{2}}{\omega ^{2}}  \nonumber \\
\therefore B &=&\frac{3}{2}\frac{N_{s}N_{v}v_{F}k_{F}}{\pi \hbar }\frac{\Omega _{cell}}{%
a^{2}}  \label{Formula_for_B_graphene}
\end{eqnarray}%
were the graphene  direct cell area $\Omega_{cell} = \sqrt{3}a^2/2$  is needed for conversion between continuous and discrete transforms.
For graphene $N_{s}=N_{v}=2.$ (i.e. 2 spin orientations and 2 valleys,
corresponding to 2 Dirac points.) In weakly interacting electron theory, the
Bloch energy in each valley is linear, $\varepsilon _{\vec{k}}=v_{F}\left| 
\vec{k}\right| $where $\vec{k}$ is measured from each corresponding Dirac
point and is assumed small compare to a reciprocal lattice vector. \ Here
the Fermi velocity $v_{F}$ is independent of doping and is about 
$0.85 \times\ 10 ^{6}$ m/sec  . \ \  

The Fermi momentum depends on doping level. \ By counting occupied states of
each spin projection in each valley we get electrons occupying k-space Fermi
circles of radius $k_F$ and Bloch energies up to $\varepsilon _{F}=v_{F}k_{F}$ where
\[
k_{F} =2\Omega _{cell}^{-1/2}\sqrt{\frac{\pi N_{dopant}}{N_{s}N_{v}}}
 =\left( 7.\,\allowbreak 74\times 10^{9}\;m^{-1}\right) 
\sqrt{N_{dopant}}
\]
Putting these results into (\ref{Formula_for_B_graphene}) we have%
\begin{equation}
B = 1.\,\allowbreak 03\times 10^{50}\sqrt{N_{dopant}}%
\;\;Joule^{-1}\sec ^{-2}  \label{B_for_doped_graphene_numerical}
\end{equation}
For the case of a $2\times 2$ $\ $superlattice of completely ionized Li
atoms we have the number of donated electrons per regular graphene cell =$%
N_{dopant}=1/4.$

The value of $B$ from (\ref{B_for_doped_graphene_numerical}) is used in Eqs. (\ref{Realspace_chi0_pp}) and (\ref{Realspace_chi0_ppprime}) to specify the charge response in the nn-MBD+C scheme.
\subsection{Avoiding double counting in Li@graphene}
The previous paragraph establishes the in-plane response of  dopant electrons in graphene  within nn-MBD+C. However the polarizability of these dopant $\pi_z$ electrons perpendicular to the plane is not included in the 2D Bloch response. With 1 Li  per $2\times 2$ supercell containing 8 carbon atoms, each  carbon atom has an extra electron 1/8 of the time, producing an anion. We therefore modified the carbon atom polarizability in the MBD part of the nn-MBD+C theory by partially using the carbon anion polarizability as follows:
  \begin{eqnarray}
  \alpha_{xx}(\omega)=\alpha_{yy}(\omega) = \alpha^C(\omega) \label{Alpha_xx_MBD_DopedGr}\\
  \alpha_{zz}(\omega) =(7/8) \alpha^{C}(\omega) + (1/8) \alpha^{C-}(\omega) \label{Alpha_zzz_MBD+C_DopedGr}
  \end{eqnarray}
  Here $\alpha^{C}$ and $\alpha^{C-}$ are the (isotropic) polarizabilities of the carbon atom and carbon anion C- .  The  $x$   and $y$ directions lie in the plane, and $z$ is perpendicular. We took these polarizabilities and associated oscillator frequencies from Bucko and Gould \cite{bucko-gould}.  
 In Figs \ref{Log_plot_energies_Li_gaphene}  and \ref{Linear_plot_energies_Li_graphene} the nn-MBD+C  dispersion energy is plotted vs. distance $D$ between the planes of C atoms. (dashed red line).

\subsection{FI-MBD for fully ionized Li@graphene}
\begin{figure}[tbp]
\includegraphics[width=0.9\textwidth]{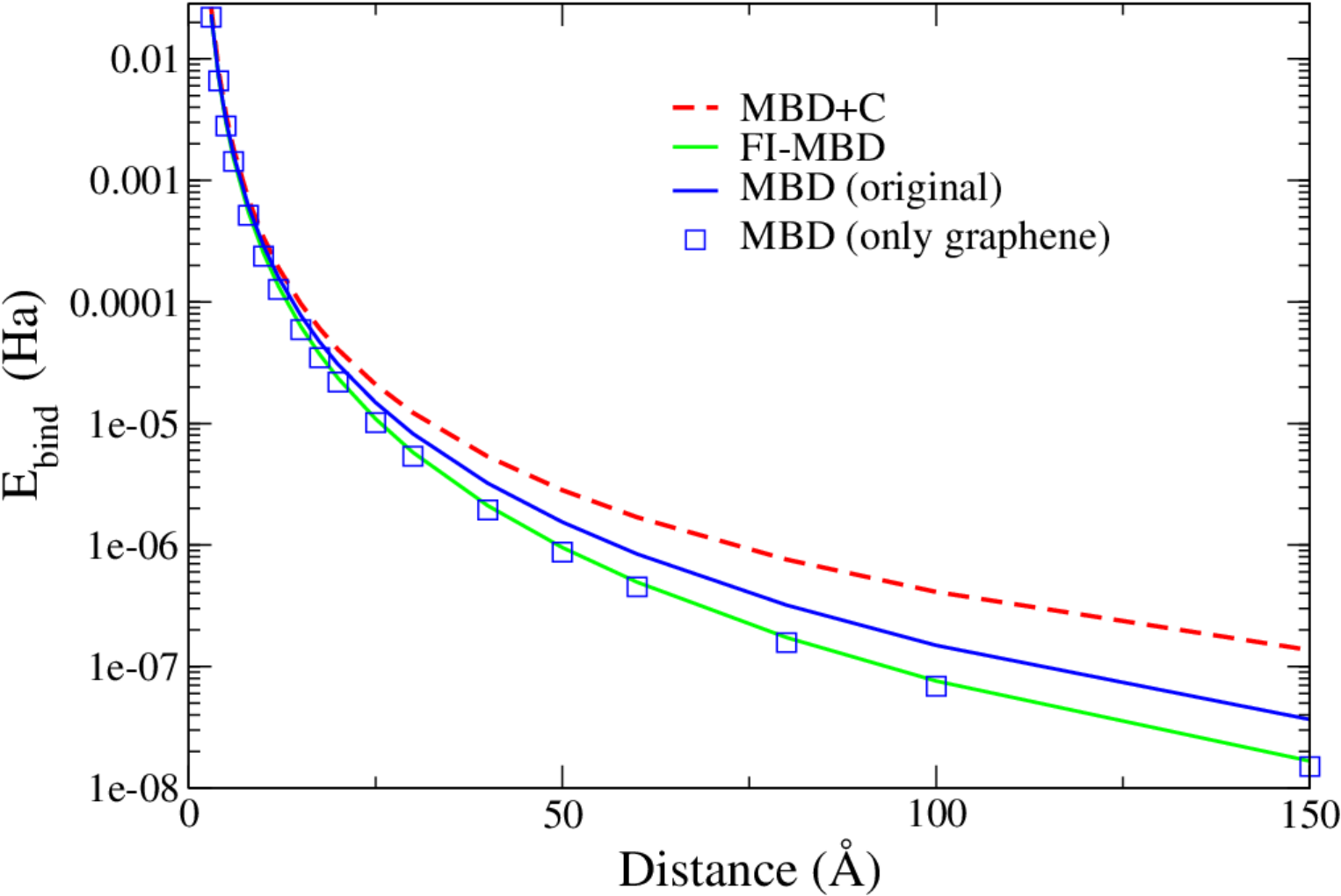}
\caption{
Dispersion energy between two parallel graphene sheets decorated on their outside surfaces by a $2\times 2$ array of lithium dopant atoms. The separation distance  $D$ is measured between the planes containing the carbon nuclei.  Results from the following MBD-type theories are potted:  the new nn-MBD+C scheme; original MBD; FI-MBD; original MBD  without the Li dopants.   The nn-MBD+C results are dominant at large distances, as expected from analytic asymptotic theory.  
\label{Log_plot_energies_Li_gaphene}
}
\end{figure}

\begin{figure}[tbp]
\includegraphics[width=0.9\textwidth]{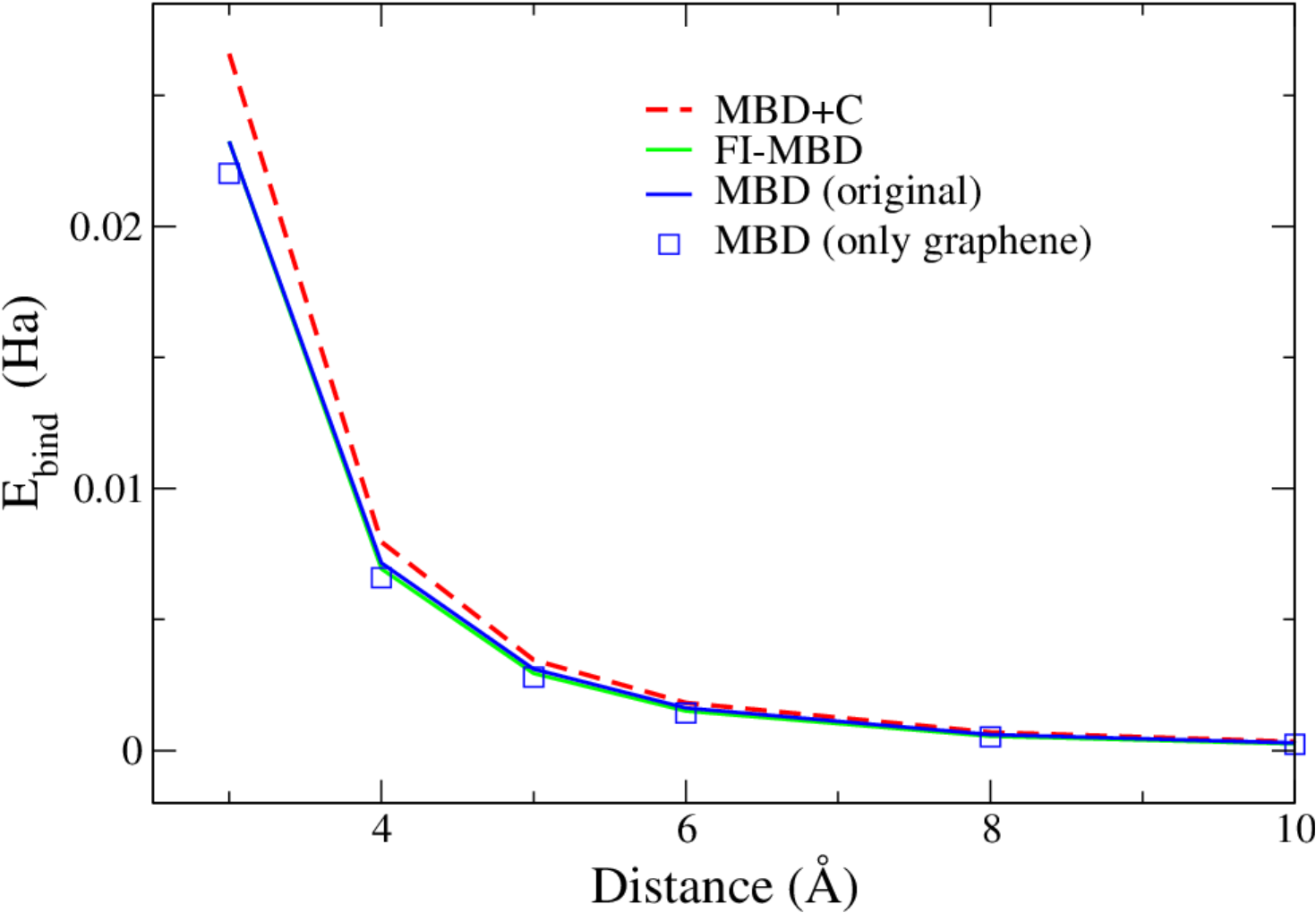}
\caption{Dispersion energy between two parallel graphene sheets decorated on their outside surfaces by a $2\times 2$ array of lithium dopant atoms, showing detail for smaller distances down to near-contact. The separation dsatance  $D$ is measured between the planes containing the carbon nuclei.  Results from the following MBD-type theories are potted:  the new nn-MBD+C scheme; original MBD; FI-MBD; original MBD  without the Li dopants.   In contrast to the asymptotic results shown in the previous Figure, here the nn-MBD+C dispersion energy no longer dominates, but near to contact at $D=3$ Angstrom it is 
$15\%$ \ larger than the MBD energy.
\label{Linear_plot_energies_Li_graphene}
}
\end{figure}

We also  used a scheme of fractionally ionic type (FI-MBD) \cite{bucko-gould} to investigate the fully ionized case.  Here Li was  treated as a  cation and the C atoms were assumed to have an isotropic polarizability, being C atoms 7/8 of the time amd  C- anions 1/8 of the time to account for the dopant electrons:.
\begin{equation}
\alpha(\omega) =(7/8)\alpha^{C}(\omega) + (1/8)\alpha^{C-}(\omega)
 \label{Alpha__FI-MBD_DopedGr}
\end{equation}
 With this  choice of polarizability we used regular MBD to calculate the inter-layer dispersion energy.
This differs from MBD+C (Eq.  (\ref {Alpha_zzz_MBD+C_DopedGr})) in that the polarizability of the dopant electrons is here described as anionic, for polarizability in all three space directions, thereby missing gapless MBD+C-type metallic electronic response in the $xy$ plane.
 In Figs \ref{Log_plot_energies_Li_gaphene} and \ref{Linear_plot_energies_Li_graphene} the FI-MBD  dispersion energy is plotted vs. distance $D$ between the planes of C atoms (pale green solid line).  The FI-MBD result agrees closely with the energy for original MBD  for graphene sheets without Li (open squares),  a fact that we attribute to the reduction of the C$^-$ polarizability due to the reduced volume based on Hirshfeld analysis. .

\begin{figure}[tbp]
\includegraphics[width=0.9\textwidth]{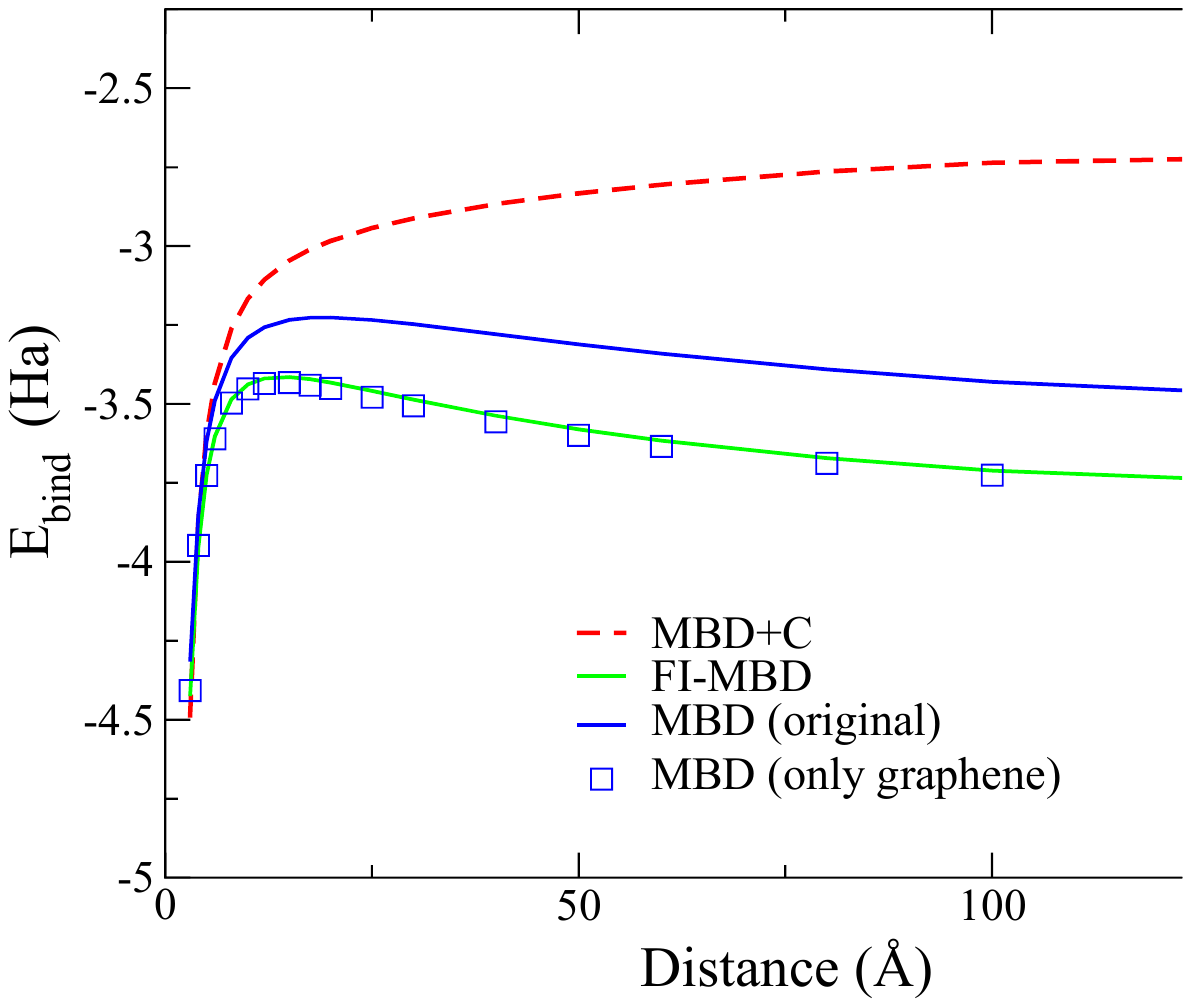}
\caption{
Effective power decay exponent $p$ of the dispersion interaction between Li-decorated graphene sheets.  Details are as for the previous two figures.  Analytic theory  predicts asymptotic values  
$p=-2,5$ for 2D metals and $p=-4$ for 2D insulators as  $D \rightarrow \infty$.
\label{Decay_exponent_p_Li_graphene}
}
\end{figure}

\subsection{Comparison between nn-MBD+C, MBD  and FI-MBD for Li-doped graphene}
Above we investigated the dispersion   energy  between two graphene sheets that  are decorated  on their outer surfaces with a $2\times2$ superlattice of Li atoms .  We implemented three theories:
\begin{enumerate}
\item{the new nn-MBD+C theory with fully ionized Li}
\item{ the original MBD, necessarily with non-ionized Li; and} 
\item{FI-MBD with fully  ionized Li and therefore partially ionized carbon.}
\end{enumerate}

Figures \ref{Log_plot_energies_Li_gaphene},  \ref{Linear_plot_energies_Li_graphene}  and   \ref{Decay_exponent_p_Li_graphene} show the results. At asymptotic separations ($D> 50$ Angstrom) the nn-MBD+C energy strongly dominated the energy from the other two theories  (Fig \ref{Log_plot_energies_Li_gaphene}), tending towards the correct power law decay exponent $p=-2.5$ known for 2D metals from asymptotic analytic RPA theory (Fig.  \ref{Decay_exponent_p_Li_graphene}) \cite{dobson-white-rubio, bostrom-sernelius}.  This was of course expected because only MBD+C recognizes the 2D metallic behavior in doped graphene.

 However we have nn-MBD+C results all the way in to near-contact at a separation of $D=3$ Angstrom.  It was not known quantitatively till now how well MBD-class theories would perform for 2D metals at such small separations, though some estimates  \cite{dobson-es} have suggested that a correction would be needed for Type-C physics for doped graphenes.
In fact at at $D=3$ Angstrom we find here that nn-MBD+C gives 15\% more attractive dispersion energy than original MBD.  FI-MD agrees almost exactly with original MBD, despite the fact that FI-MBD explicitly considers the ionized character of lithium and carbon.  The similarity of MBD and FI-MBD here might seem surprising since the  C- anion encountered in F-MBD is more polarizable than the C atom used in MBD.  The explanation may lie in the Hirshfeld atomic-volume scaling that we applied to polarizabilities  in all types of MBD.

In summary, the results in this Section  show that a theory such as MBD+C is needed in order to calculate the dispersion energy of this 2D metal sandwich at all separations, even near to contact.

\section{Dispersion interaction between parallel armchair carbon nanotubes}
\subsection{CNT geometry and electronic bandstructure}
Carbon nanotubes (CNTs) are cylindrical structures that can be visualized by rolling up a rectangular section of planar graphene. See the book by Saito and Dresselhaus \cite{Saito_Dresselhaus_book}. The shape of the rectangle is specified by integer  folding indices (m,n). Armchair carbon nanotubes have folding indices of form  (n,n) and they are intrinsically conducting along the tube axis but insulating in the circumferential direction, as explained qualitatively below.
 The direct unit cell of an (n,n) tube contains 4n carbon atoms arranged around the circumference of the tube  in an armchair texture.  Taking the z (axial) direction to be vertical for definiteness,  we note that the atoms within a cell are arranged in an upper ring  or 2n atoms forming the arms of the armchairs, and a lower ring of 2n atoms forming the seats of the armchairs.  Within a ring the atoms are clumped into pairs.  All atoms in a given ring have the same z coordinate.  We label each  ring  with an integer $I$,  $-\infty < I < \infty$. 
 There is no energy gap between occupied and unoccupied Bloch state with different crystal momentum $k_z$  parallel to the axis. Thus electronic charge can freely move in the z direction  between rings. However the  Bloch states for motion around the circumference  are discrete because of periodic boundary conditions, and are full up to a gap. As a result any excess  charge within a ring has a frozen spatial profile with an equal amount of charge on each atom of the ring, a conclusion supported by detailed bandstructure calculations.   This means that the tube can be regarded as a one-dimensional conductor similar to the gold atom chain studied earlier in the present wok.  However the charged objects moving one-dimensionally are not not electrons localized on a single atom, but are grainy rings containing one electronic charge equally distributed across 2n atoms.
The one-dimensional electronic Bloch bandstructure of (n,n) CNTs has been worked out in 
 Ref. \cite{Saito_Dresselhaus_book}.    There are two conduction bands  with two inequivalent non-avoided Dirac crossing points $\pm k_{Dirac}$ within the Brillouin zone.  The Fermi velocity $|v_F =\hbar^{-1}d\varepsilon/dk|$ is the same as in 2D graphene.
\section{One-dimensional density ( charge) response of (n,n) CNTs}
Here we introduce a simple atom-ring--based Ansatz for the electronic density response  of the
conduction electrons .  Following the equivalence to an atomic chain as pointed out in the previous section,  we propose the following "nearest-ring"  Ansatz identical in form to the nearest-neighbor Ansatz  (\ref{Chi0Ansatz)_1Dchain}) introduced for monoatomic chains:
\begin{equation}
\chi _{0IJ}^{Ansatz}\equiv \chi _{0}\left( I-J\right) =B\omega ^{-2}\left(
\delta _{IJ}-\frac{1}{2}\delta _{I,J+1}-\frac{1}{2}\delta _{I,J-1}\right)
\label{Chi0Ansatz)_(nn)CNT}
\end{equation}%
Here $I$ and $J$  now label rings  of atoms rather than individual atoms as in (\ref{Chi0Ansatz)_1Dchain}).  Because this form is identical to (\ref{Chi0Ansatz)_1Dchain}) we can determine the constant $B$ to reproduce the long-wavelength  charge response of 1D Bloch electrons in the same way, obtaining, as in (\ref{Bfor1DCase}):
\begin{equation}
B=\frac{2N_s N_v v_{F}}{\pi \hbar R}
\end{equation}%
Here $ R=a/2 =1.23 $ Angstrom  is the spacing in the z (axial) direction  between rings of atoms in the tube.
While the Kohn-Sham response has one-dimensional form, the Coulomb interactions among charges and dipoles have to be calculated recognizing that the charges are distributed equally around the $2n$ atoms in a ring.
\subsection{Avoiding double counting  in nn-MBD+C for (n,n) CNTs}
As always, we need to ensure we do not double-count the response in nn-MBD+C. 
There are 2 1D conduction bands in an (n,n) nanotube.  Each band is half full of spin-up electrons and half-full of spin-down electrons. General  Bloch band theory shows that there is one Bloch orbital in each band, per direct unit cell. Thus there are 2 conduction electrons per cell, and each cell contains $4n$ atoms. Hence  there are $2/(4n) = 1/(2n)$ conduction electrons per carbon atom.  These conduction electrons must not be allowed to contribute to the atomic polarizability in the z direction, as their response is already accounted for in nn-MBD+C by the conduction electron charge response  $\chi_0$.  A given atom donates an electron to the conduction bands (and should therefore have a cationic polarizability), with probability $f=1/(2n)$.  
For the polarizability response in nn- MBD+C we therefore keep the full carbon atomic polarizability in the MBD part, EXCEPT that the  carbon polarizability component in the z direction (parallel to the tube axis) should be reduced as follows:
\begin{eqnarray}
\alpha _{xx} &=&\alpha _{yyy}=\alpha ^{C} \\
\alpha _{zz} &=&\frac{2n-1}{2n}\alpha ^{C}+\frac{1}{2n}\alpha ^{C+}
\label{Alpha_zz_nnTube}
\end{eqnarray}
Here the z axis is the nanotube axis. $\alpha ^{C}$ is the polarizability of a
neutral carbon atom and $\alpha ^{C+}$ is the polarizability of the carbon 
cation $C^{+}$.
\subsection{Numerical implementation of nn-MBD+C for parallel (4,4) carbon nanotubes }
In all of the calculations reported below for this system,  just as for the
 doped graphene  calculations reported above, we implented nn-MBD+C using Gaussian Coulomb damping and Hirshfeld-volume-based polarizability renormalization. 
We  considered "on-top" registry where atoms in the two tubes were directly opposite.

\begin{figure}[tbp]
\includegraphics[width=0.9\textwidth]{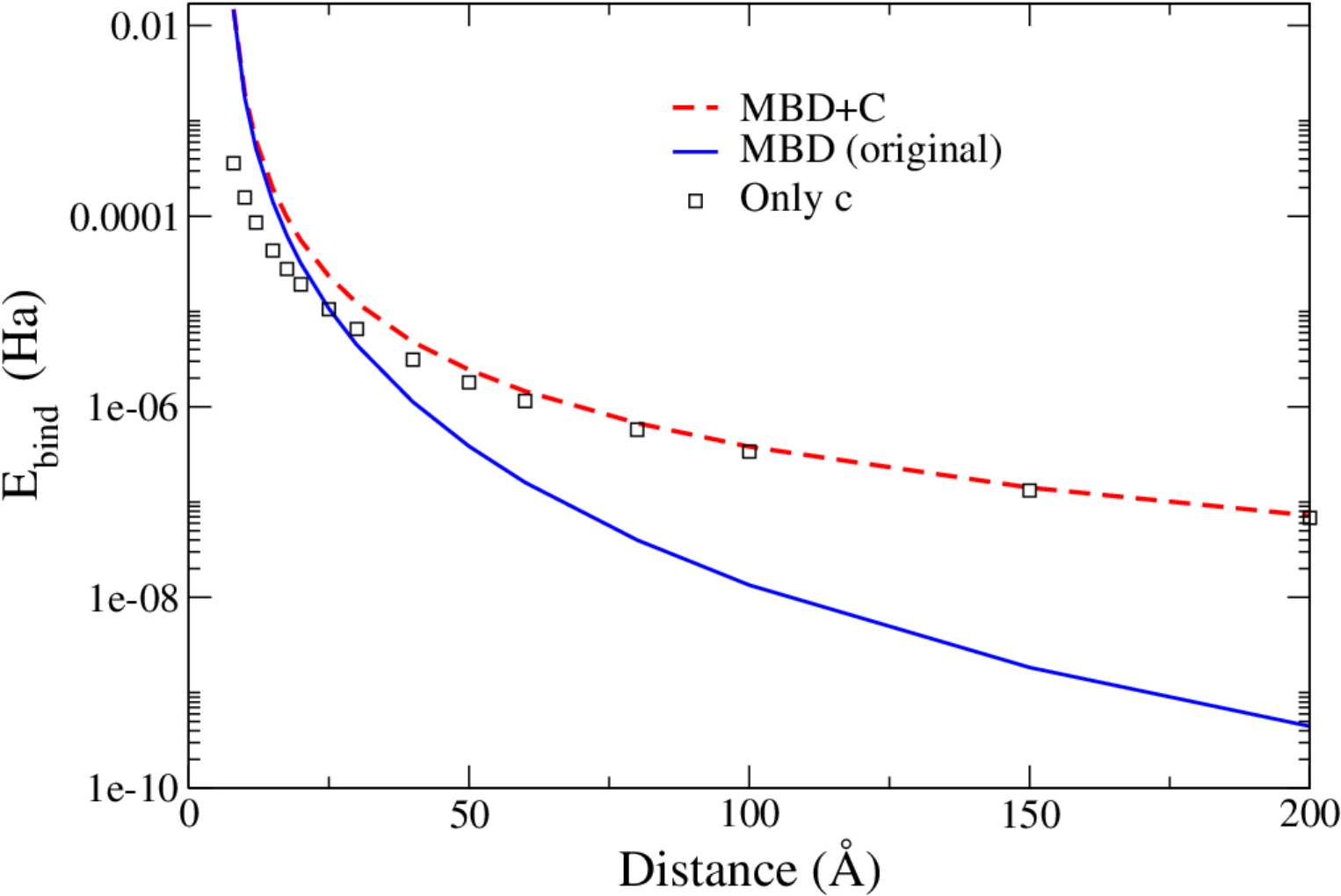}
\caption{
Dispersion energy between parallel conducting (4,4) carbon nanotubes in nn-MBD+C and original MBD theories.  Here the separation distance $D$ is measured between the central axes of the tubes. The nn-MBD+C energy dominates in the asymptotic regime of large separations $D$ .
\label{Energy_log_44CNTs} 
}
\end{figure}

\begin{figure}[tbp]
\includegraphics[width=0.9\textwidth]{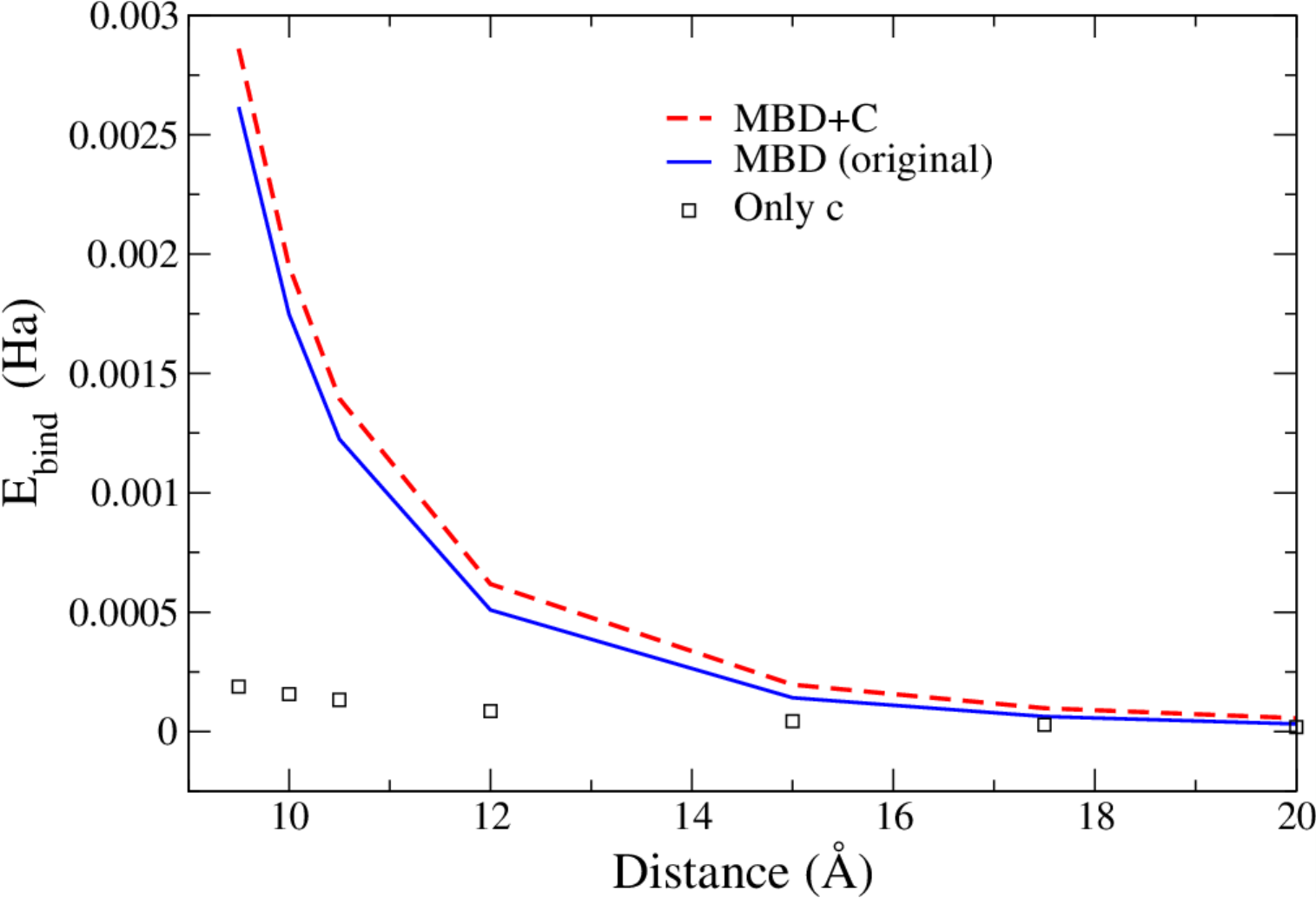}
\caption{
Dispersion energy between parallel conducting (4,4) carbon nanotubes in nn-MBD+C and original MBD theories, at smaller inter-axis separations $D$. 
\label{Energy_44CNTs} 
}
\end{figure}

\begin{figure}[tbp]
\includegraphics[width=0.9\textwidth]{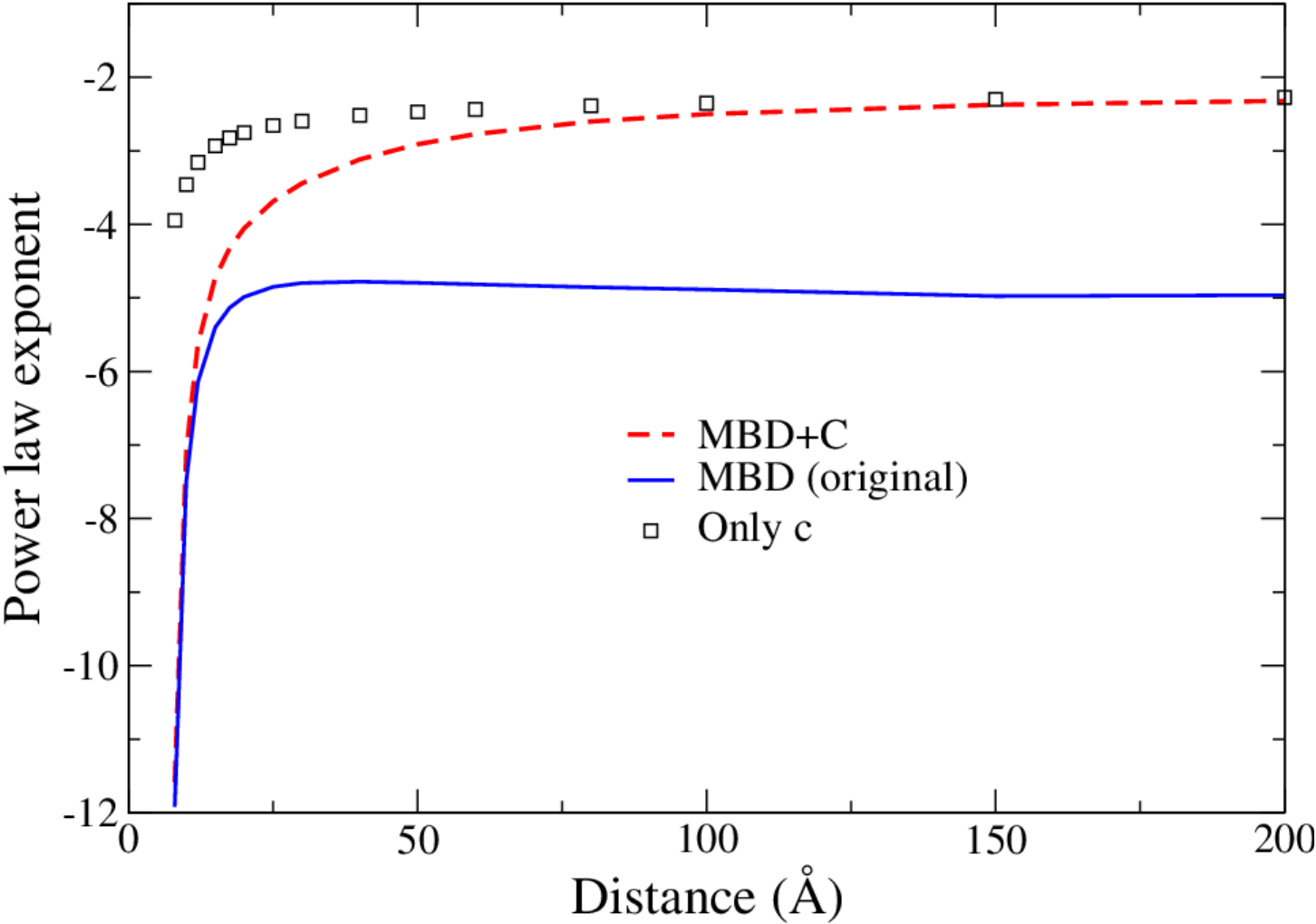}

\caption{
Effective power law decay exponent 
 $p=d \ln  |E\left( D\right)| /d (\ln D)$
of the dispersion interaction between two parallel (4,4) carbon nanotubes, 
 from nn-MBD+C and MBD theories.  The expected asymptotic decay laws $p \approx -2+$ and $p=-5$ are found for nn-MBD+C and MBD respectively.
\label{Eff_power_44CNTs} 
}
\end{figure}

The dispersion energy results are given in Figs. \ref{Energy_log_44CNTs} and \ref{Energy_44CNTs}, and the power law decay is analyzed in Fig. \ref{Eff_power_44CNTs}.  

 Again the numerical  results from nn-MBD+C approached the known asymptotic decay law for 1D  metals at large separations, and the nn-MBD+C energies dominated those from the original MBD in this asymptotic regime.  

Near to contact nn-MBD+C gives a dispersion energy  9\%  greater than MBD, and the percentage difference grows as the separation $D$ increases.

\section{Summary}

We have proposed a method, here termed "MBD+C", for introducing metallic character into
discrete atom-based dispersion energy theories such as MBD\cite{mbd,mbdrs}. This
allows the theory to treat both non-metals and metals in a natural way, accounting for the ''Type-C van der Waals non-additivity'' \cite{dobson-abc,dobson-book} that  strongly  affects the asymptotic dispersion interaction between low-dimensional metals.
 This contrasts with current theories such as
MBD, FI-MBD \cite{Gould_FIMBD_2016} and UMBD \cite{Kim_uMBD_2020}, which miss the  Type-C physics.
  
The essential new  feature is to allow each ''atom'' to have a transient
electric charge that can move between atoms, in addition to the transient
dipole allowed in MBD. \ In our MBD+C theory this process is described by a discrete form of the
independent-electron density (charge) response function : $\chi _{0}\left( \vec{r},%
\vec{r}^{\prime },\omega \right) \rightarrow \chi _{0IJ}\left( \omega
\right) $ \ where $I$ and $J$ label ''atoms'' located at positions $\vec{R}_{I}$, $\vec{R}_{J}$.
 In MBD+C the response function$\chi_{0IJ}$  is used in
time-dependent Hartree (direct Random Phase Approximation, dRPA) equations 
(\ref{Gen_MBD+C_for_p},\ref{Gen_MBD+C_for_n}) that   
couple \ all dynamic dipoles and charges via their \ Coulomb interaction. \ \
This yields coupled collective charge+dipole modes with frequencies
 $\omega_{i}$. The sum of mode zero-point energies \ \ $\sum_{i}\hbar \omega _{i}/2$
gives a correlation energy that contains the dispersion interaction.  
 For non-metals the charge  response function
$\chi_{0IJ}$ vanishes and the  theory becomes identical with regular MBD, if we choose a harmonic oscillator model for the atomic polarizabilities.
 
The response $\chi _{0IJ\text{ }}$can be derived
microscopically via a tight binding approach, and in large metallic systems it is
known to have an algebraically decaying tail as
 $|\vec{R}_{I}-\vec{R}_{J}| \rightarrow \infty$.  Nevertheless we showed that a simple near-neighbor
Ansatz for $\chi _{0IJ}$ $\ $leads to the correct dispersion relation for
long-wavelength plasmons, and this is the essential physics leading to
Type-C dispersion interactions between low-dimensional metals. This Ansatz is
described for particular cases in Eqs. (\ref{Chi0Ansatz)_1Dchain}) and (\ref%
{Chi0Ansatz_graphene}), and in more general form in Eq (\ref%
{GeneralNN_Chi0Ansatz}). The rest of the paper then uses this simplified near-neigbor  version
of our theory, which we call   "nn-MBD+C".

We first applied the nn-MBD+C theory to  a single metallic chain of atoms, where we obtained the collective charge+dipole mode frequencies 
$\omega_J \equiv \omega(q)$ analytically for each wavenumber $q$, permitting explicit verification of mode stability, which has sometimes been am issue in MBD when describing metals.  The gapless quasi-acoustic one-dimensional plasmon mode was present in nn-MBD+C  (see Figs.\ref{Fig1UncoupledFreqs} and \ref{CpledModes1Chain}), in contrast to MBD, which only has gapped polarization modes. 
We also obtained analytic expressions (Eq. (\ref{Coupled_2_chain_frequs})) for the coupled mode frequencies of  two parallel atomic metal chains for a strictly one-dimensional toy model where the atomic polarizability was non-zero only in the direction along the chain.  A one-dimensional $q$ integration as per Eq. (\ref{EvdW_from_mode_frequs}) them gave the dispersion energy.  We verified explicitly that the nn-MBD+C theory gave the correct asymptotic power law decay in both metallic and non-metallic cases:   see Figs.  \ref{EvdWNoCond}, \ref{EvdWNoPol}  and  \ref{PowerExponentStrbictly1D} .  For near-contact separations,  nn-MBD+C gave results similar to regular MBD: see Fig.   \ref{EvdWStrictly1D}. 

Following these successful tests on the above toy 1D model,  we  numerically evaluated the nn-MBD+C dispersion  energy for three low-dimensional metals:
\begin{enumerate}
\item{two parallel monoatomic chains of equally spaced gold atoms: see Figs.  \ref{LogEnergy_2Au_chains}
 - \ref {Eff_power_2Au_chains}}
\item{ two parallel sheets of metallic lithium- doped graphene: see Figs.   \ref{Log_plot_energies_Li_gaphene}
 - \ref{Decay_exponent_p_Li_graphene}}
\item{two parallel conducting (4,4) carbon nanotubes: see Figs.  \ref{Energy_log_44CNTs}  - \ref{Eff_power_44CNTs}}
\end{enumerate}

In these calculations, nn-MBD+C seamlessly described the dispersion energy for configurations ranging from near-contact through intermediate spacings to the asymptotic regime of large separations.   We did not treat the full-contact regime of  strong inter-metallic bonding, as this is already well described in semi-local density functional theory: the nonlocal inter-atomic correlation energy described by MBD-type theories is of marginal relevance there. On the basis of our near-contact calculations, we expect that Type-C physics will affect the large contact energies by less than 1\%.  The regimes that we did cover are relevant  for self-assembly, docking, catalysis,  reaction barrier height. and molecular dynamics for example.

nn-MBD+C describes the  dispersion interaction seamlessly from near-contact through to the asymptotic regime.  As expected, in all systems tested, the asymptotic nn-MBD+C dispersion energy dominated that from existing  MBD-type theories, with numerical results confirming analytic predictions for the  exponent $p$ in a type-C asymptotic power-law decay 
$E \propto D^p$.  Till now it has not been clear whether metallic Type-C effects can be significant outside the asymptotic regime, and we have now addressed this issue using MBD+C.   

As already noted, at full metallic contact the large binding energy is insensitive to Type C effects. However the present work finds that,  just outside contact,  the nn-MBD+C dispersion energy  is modestly greater than that from regular MBD. At separations 1/2 Angstrom outside metallic  contact, where the dispersion energy already dominates the attraction, we found that the nn-MBD+C energy is greater than the MBD energy by 4\%, 11\%  and 9\%  in the cases studied so  far.    This confirms that, while regular MBD gives a reasonable description of low-dimensional metals  in the near-contact region, it is not necessarily quantitatively reliable there.  Furthermore  the percentage error  increases with increasing separation.  It is therefore unsafe to ignore Type-C effects on the dispersion energy between low-dimensional metals, in any regime of inter-system separation except full metallic binding at contact.   MBD+C provides an efficient way to include these Type-C effects.

The nn-MBD+C method is not significantly more demanding computationally than regular MBD, requiring operations with $4N\times 4N$ matrices compared with $3N\times 3N$ for regular MBD,  where $N$ is the number of atoms . In infinite periodic systems a wavevector analysis permits finite matrices to be used, and the MBD+C matrices are not significantly larger than those in regular  MBD.

The present near-neighbor form of  MBD+C, while numerically efficient, is not yet a "black box" in the sense that it requires some user insight about the  bandstructure of the metallic system, in order to fix the form of the near-neighbor Ansatz for the charge density response in Eqs.  (\ref{Chi0Ansatz)_1Dchain}) -  (\ref{GeneralNN_Chi0Ansatz}).  The examples given here show how to do this for the most important cases of Type-C physics, namely low-dimensional metals of macroscopic length, or structures that contain these.   However, tight-binding(TB)  theory is known to give a reasonable account of metallic Bloch bandstructure (see the book by Saito and Dresselhaus  \cite{Saito_Dresselhaus_book} for the case of metallic graphenics, for example). The full form of MBD+C will obtain the metallic charge density  response $\chi_{0IJ }$  from a TB approximation for the global orbitals, without the need for a  near-neighbor approximation for the response.  Here the inputs will be TB inter-atom Hamiltonian hopping matrix elements $t_{IJ}$ from the literature, without the need for user insight.  This approach could  provide a black-box theory.

\section{Supporting Information}
Kohn-Sham electronic density response of  crystalline one- and two-dimensional metals.

\section{Artwork for Table of Contents}
\includegraphics[width=0.9\textwidth]{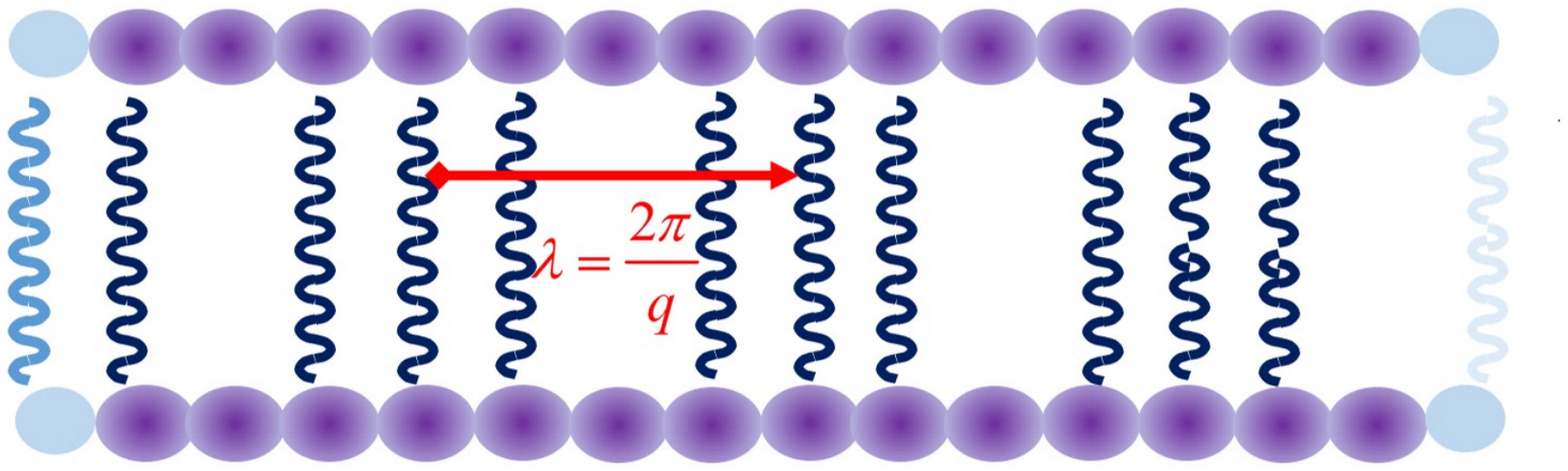}

\begin{acknowledgement}
We thank Tim Gould for useful discussions.
\end{acknowledgement}


\bibliographystyle{achemso}
\bibliography{MBD+C_V11}

\end{document}


\begin{abstract}
\end{abstract}

\section{Density response of  independent Bloch electrons in a 1D metal}

The one-dimensional (1D) system is taken to consist  of $N \rightarrow \infty$  unit cells each of length $R$ so that the total length is  $L=NR  \rightarrow \infty$. 
We treat the electrons in the 1D conductor as independent but subject to a 
 grouandstate Kohn-Sham (KS) potential that is periodic, $V(x+R)=V(x)$. 
 The KS eigenstates 
 $\psi_{nk}(x)$ are labeleed by a crystal momentuum $k$ such that 
 $\psi_{nk}(x+R)=exp(ikR)\psi_{nk}(x)$,
and by an integer  band index $n$. A full band carries no current.   We will asume that we are dealing with a partially occupied conducton band, and will drop the index $n$ from here on.
Under periodic boundary conditions across the entire length, the allowed crystal momenta  are labelled by an integer $i$:
\[
k_{i}=i\frac{2\pi }{L},\,\,-N/2<i<N/2
\]

The number of allowed $k$ values in an interval $\Delta k$ is then
\begin{equation}
\Delta N = L \Delta k /(2\pi)
\label{Delta_N}
\end{equation}
 For simplicity we assume inversion symmetry so that the KS energy eigenvalue 
$\varepsilon(k)$  is a function of $|k-k_0|$ for some $k_0$. We also assume that for the "pocket"  of  condution electrons under consideration  
$\varepsilon (k)$ increases monotonically with $|k-k_0|$ up to the highest occupied energy state.   These assumptions are widely valid - e.g. for  free electrons in 1D,
$k_0=0$ and $\varepsilon (k) = \hbar^2  k^2 / (2m)$. 
  For electrons in the monatomic  gold nanowire studied in the main paper, $k_0=0$ and the monotinic condition is satisfied (See Fig. 4 of Jariwala et a.l  \cite{jariwala} where the point $k=0$ is labelled $\Gamma$) . For a metallic (n,n)  carbon nanotube  there are two Dirac points $k_{01}$, $k_{02}$
  and $\varepsilon (k) \approx v_F |k-k_{0i}| $ for electrons in the conducting pocket near each Dirac point $k_{0i}$. \cite{Saito:98} 
   
 The Aufbau principle  implies that, in the unperturbed ground  state, the Bloch orbitals are filled up to some Ferni energy $\varepsilon_F$.  The above symmetry and monotonicity assumptions mean that the states are then filled up to a  Fermi momentum $k_F$    where   $-k_F < k-k_0 < k_F$,  Orbitals in each pocket are occupied by electrons of  $N_{s}$ possible spin projections (usually $N_s=2$).

We consider the response of the Bloch electrons to a small,  slowly-varying, long-wavelength external potential  $ \Delta V (x,t)$ where
\begin{equation}
\Delta V (x,t) = -e\phi(x,t) = \Delta V exp(i(qx-\omega t)), qR<<1,  v_Fq <<  \omega << \varepsilon_F /\hbar
\label{DrivingPotential}
\end{equation}

 We use the semi-classical approximation (SCA)  \cite{Ashcroft:Book:76}, which is valid in this limit. The electron momentumn and velocity are ill-defined in a Bloch  orbital because they change constantly under the action of the periodic crystal potential.  In the SCA it is noted that the expectation of velocity in a Bloch state is

\[ 
v = \hbar^{-1} \partial \varepsilon  / \partial k 
\]
\\

where the Bloch state extends over the entire chain of length $L$. 
 Note that $v$  is also the group velocity - i.e. the velocity of the envelope of  a spatially localized wavepacket made by superposing Bloch states with a spread of crystal momenta centered on the value $k$.

The veocity determines charge transport, and the expectation of the  charge current  $j$ in am extended  Bloch state is 

\begin{equation} 
j  =\frac{-e}{L}v = \frac{-e}{L \hbar} \partial \varepsilon  / \partial k 
\label{BlochCurrent}
\end{equation}

Under the action of the  external potential energy (\ref{DrivingPotential}) ,  in the SCA  the crystal momentum  $k$ of each Bloch orbital changes in response to the applied force $F$ according to the deceptively classical-appearing formula
\[
\hbar \frac{dk}{dt}= F = -iq \Delta V exp(i(qx-\omega t),    q \rightarrow 0
\]
 Each wavevector will thus  be of the form 
\[
k_{i}^{\prime }=k_{i}+\Delta k\exp \left( -i\omega t\right) 
\]
where 
\begin{eqnarray}
-i\omega \hbar \Delta k &=&-iq   \Delta V\  \\
\Delta k &=& \frac{q \Delta V}{\hbar \omega }
\label{Delta_k}
\end{eqnarray}
In the unperturbed condition the currents from all the occupied states
cancel out.   In the excited  state the occupied orbitals have cystal momenta $k$ lyng in the range

\[
-k_F +\Delta k < k-k_0  < k_F + \Delta k
\]

Thu net current comes from from a group of electrons numbering
 $\Delta N =(L/2\pi) \Delta k$ with crystal momentum $k_0+k_F$ and  velocity
 $v_F =\partial \varepsilon^{(0)} \partial k |_{k_F}$ , minus the contribution of the same number of previously-occupied states with $k=k_0-k_F $ and velocity $-v_F$.

The total current is thus

\begin{equation}
j=\Delta N \frac{-e}{L}(v_F - (-v_F)) =\frac{-e}{2\pi}\frac{q\Delta V}{\hbar \omega}2v_F
\label{j}
\end{equation}
where (\ref{Delta_N}) and (\ref{Delta_k}) were used in the last step.

This flow of current gives rise to a charge density perturbation
 $\Delta \rho (x,t)=-e\Delta n \exp(i(qx-\omega t))$ that must satisfy the continuity equation 
$0=\partial \Delta \rho / \partial t + \partial  j /\partial x$  (charge conservation) so that 
\begin{eqnarray}
0 &=& -i\omega (-e\Delta n)+iq j  \\
\Delta n &=& -\frac{q}{e \omega} j
\label{Delta_n_2}
 \end{eqnarray}

Using (\ref{j}) we can write (\ref{Delta_n_2}) in the form
\[   
\Delta n =  \chi_0 (q,\omega) \Delta V 
\]
where the Kohn-sham (independent-
electron) density response function for 1D Bloch electrons  for $qR<<1,  v_Fq <<  \omega << \varepsilon_F /\hbar$ is

\begin{equation} 
 \chi_0 (q,\omega) = \frac{q^2 v_F}{\pi \hbar \omega^2}
\label{Chi0_1pocket}
\end{equation}
 Eq  (\ref{Chi0_1pocket}) applies to each  "poclet"  of conducting elctrons.  The pockets can differ by the spin projection of the electrons (e.g. up or down).  Where a Bloch band structure $\varepsilon(k)$ has multiple valleys that cross the Fermi energy, as for example in some graphenic conductors, this can also  lead to multiple pockets of conduction electrons.  If there are several  conducting pockets then
\begin{equation} 
\chi_0(q,\omega) = \frac{q^2}{\pi \hbar \omega^2} \sum_P  v_F^{(P)}
\end{equation}
where $P$ labels the various pockets. If there are $N_s$ allowed spin orientations and $N_v $ inequivalent  valleys then the number of pockets equals $N_s  N_v$ .  If all pockets  have the same Fermi velocity, $v_F^{(P)}=v_F$,  then

\begin{equation}
 \chi_0 (q,\omega) =N_s N_v  \frac{v_F q^2}{\pi \hbar \omega^2}
\label{Chi0Final}
\end{equation} 

For  unmagnetized metallic systems, both spin-up and spin-down bands are partially occupied so that $N_s = 2$. For a chain of single monovalent metal atoms only one valley crosses the Fermi energy so that $N_v=1$.  For metallic  (n,n) carbon nanotubes there are valleys centered on two inequivalent Dirac ponts at the Fermi energy  \cite{Saito:98} so $N_v= 2$.

Equation(\ref{Chi0Final}) is quoted as Eq (5) in the main manuscript.

As a check we can evaluate (\ref{Chi0Final}) for free electrons of mass $m$, for which 
$v_F = \hbar k_F m^{-1} $  and $N_v = 1$. The occupied orbitals have wavenumbers in the range $-k_F <k< k_F$ and each orbital is occupied by one spin-up and one spin-down electron (i.e. $N_s =2$).    The total number of electrons $N$ is then

\begin{equation}{
N = N_s \frac{L}{2\pi}(2k_F) \;\;\;  \text so} \;\;\;  k_F = \frac{\pi N}{L N_s}
\end{equation}

Then (\ref{Chi0Final}) becomes
\begin{eqnarray*}
 \chi_0 (q,\omega) &=& N_s N_v  \frac{v_F q^2}{\pi \hbar \omega^2} \\
 &=&N_s \cdot 1 \cdot \frac{\hbar  (\pi N)/(L N_s) )m^{-1} q^2}{\pi \hbar \omega^2} \\
&=& \frac {n q^2}{m \omega^2}
\end{eqnarray*}
where $n=N/L$ is the total number of electrons per unit length.   The last equation is the standard result for free electrons, valid for $q<<k_F$    and $\omega >>  v_F q$.  Eq  (\ref{Chi0Final}) above  is its generalization to Bloch electrons.

\section{Density response of  independent Bloch electrons in a 2D metal} 

For Bloch electrons in 2D with an isotropic Bloch energy $\varepsilon(|\vec{k}|)$ we can use a similar argument to the 1D argument above, obtaining

\begin{equation}
\chi _{0}^{2D}\left( \vec{q},\omega \right) \approx \frac{N_sN_v v_{F}k_{F}}
{4\pi\hbar }{}\frac{q^{2}}{\omega ^{2}}\;\;as\,q\rightarrow 0
\label{Chi0_2Dmetal}
\end{equation}
which is Eq. (6) of the main text. 
 We will not derive this result in detail here, but will check it for the free 2D unpolarized electron gas. For ths case $N_v=1$, $N_s = 2$ and 
$v_F = \hbar k_F /m$.  The occupied orbitals of each spin species fill a circle in $k$ space of radus  $k_F$.  Thus the total number of electrons is 
\[
N = 2 \frac{A}{(2\pi)^2 }(\pi k_F^2 ) \:\: so \:\: k_F^2 =2 \pi N/A 
\]
and then  (\ref{Chi0_2Dmetal}) becomes 
\[
\chi _{0}^{2D}\left( \vec{q},\omega \right) \approx 
\frac{(2)(1)(\hbar k_F /m)k_F}
{4\pi\hbar }
\frac{q^{2}}{\omega ^{2}} %
=\frac{k_F^2}{2m\pi}\frac{q^2}{\omega^2}
=\frac{nq^2}{m\omega^2}
\]
where $n=N/A$ is the total number of conduction electrons per unit area.  The last formula is the correct result for free electrons in 2D when $q<<k_F$  and $v_Fq <\omega << \varepsilon _F$: it follows from $F=ma$ plus number conservation.

\bibliography{MBD+C_V10_Suppl}